\documentclass[twocolumn,showpacs,preprintnumbers,amsmath,amssymb]
{revtex4-1}

\usepackage{color}
\usepackage{graphicx}
\usepackage{dcolumn}
\usepackage{bm}

\begin{document}

\preprint{}

\title{Quark scalar, axial, and pseudoscalar charges in the Schwinger-Dyson formalism
}

\author{Nodoka~Yamanaka}
  \email{yamanaka@ruby.scphys.kyoto-u.ac.jp}
  \affiliation{Department of Physics, Graduate School of Science,
  Kyoto University, \\
  Kitashirakawa-oiwake, Sakyo, Kyoto 606-8502, Japan}
\author{Shotaro~Imai}
  \affiliation{Department of Physics, Graduate School of Science,
  Kyoto University, \\
  Kitashirakawa-oiwake, Sakyo, Kyoto 606-8502, Japan}
\author{Takahiro~M.~Doi}
  \affiliation{Department of Physics, Graduate School of Science,
  Kyoto University, \\
  Kitashirakawa-oiwake, Sakyo, Kyoto 606-8502, Japan}
\author{Hideo~Suganuma}
  \affiliation{Department of Physics, Graduate School of Science,
  Kyoto University, \\
  Kitashirakawa-oiwake, Sakyo, Kyoto 606-8502, Japan}

\date{\today}

\begin{abstract}

We calculate the scalar, axial, and pseudoscalar charges of the quark in the Schwinger-Dyson formalism of Landau gauge QCD.
It is found that the dressed quark scalar density of the valence quark is significantly enhanced against the bare quark contribution, and the result explains qualitatively the phenomenologically known value of the pion-nucleon sigma term and also that given by lattice QCD.
Moreover, we show that the Richardson's interquark potential suppresses the quark scalar density in the Higashjima-Miransky approximation.
This fact suggests that the quark scalar density is an observable that is sensitive to quark confinement. 
For the quark axial charge, we find that it is suppressed due to the gluon dynamics.
The result of the quenched analysis agrees qualitatively with the experimental data of the isovector axial coupling constant $g_A$.
We show that the suppression of the quenched axial charge is due to a mechanism similar to that of the quark tensor charge.
In the Schwinger-Dyson equation with the leading unquenching quark-loop contribution the quark axial charge is more suppressed, due to the anomaly effect.
The quark pseudoscalar density is found to be large, and is divergent as the bare quark becomes massless.
This result is in agreement with the phenomenological current algebraic analysis, and explains well the dominance of the pion-pole contribution.

\end{abstract}


\maketitle

\section{\label{sec:intro}Introduction}

The study of the nucleon structure is one of the most important methods to clarify the dynamics of quantum chromodynamics (QCD), and many experimental and theoretical researches have been done, extending across a wide range of energy scales.
The nucleon can form five types of charges, the scalar, vector, tensor, axial vector, and pseudoscalar charges, which give the leading contribution of the nucleon form factors expanded with the exchanged momentum \cite{adler}.
Particular attention is paid to the contribution of the quark to these form factors.
In experiments, observables related to the nucleon structure are generally derived from the quark operator.
The quark contribution to the nucleon charges provides important information about the nonperturbative effects of QCD, but it also serves as a probe of more microscopic interactions, such as the weak interaction with the neutrinos or with new particles beyond the standard model.
The study of these charges are thus of crucial importance in the study of QCD and particle physics.

The quark scalar density of the nucleon $\langle N | \bar q q | N \rangle$, given by the simplest Lorentz structure, is an important quantity in the nonperturbative study of the nucleon, since it gives the renormalization group invariant contribution of the bare quark, or the explicit chiral symmetry breaking, to the nucleon mass.
This also gives important information about the relativistic structure of the nucleon.
By comparing the scalar density and the vector charge, it is possible to probe how relativistic the particles are.
The isoscalar quark scalar density is known as the pion-nucleon sigma term $\sigma_{\pi N} \equiv m_q \langle N | \bar q q | N \rangle$, and many phenomenological \cite{adler,sigmatermreview,sigmatermmodel,hai-yang_cheng} and lattice QCD \cite{sigmatermlattice,sigmatermandspinlattice,bhattacharya,strangecontentlattice,isovectorscalardensitylattice} analyses have been done. 
It is also a useful input in the direct search for dark matter \cite{darkmatter}, since it gives the strength of the interaction of the nucleon with the new particles beyond the standard model, such as the neutralino in supersymmetric models \cite{mssm}.
Moreover, the quark scalar density relates the quark level contribution of the new physics beyond the standard model to the semi-leptonic nucleon level processes such as CP-odd interactions of the electric dipole moment \cite{edm,edm2} or beta decay \cite{herczeg}.

From the phenomenological analyses, although fluctuating with a relatively large uncertainty, the pion-nucleon sigma term takes a value ranging from 70$-$40 MeV \cite{adler,sigmatermmodel,hai-yang_cheng}.
Recent studies of $\sigma_{\pi N }$ in lattice QCD show values in the lower region of this range, $\sim 40$ MeV \cite{sigmatermlattice}.
If we assume that the scalar density of the quark is carried by the valence constituent quarks and that the scalar density carried by each quark contributes additively (this assumption relies on the fact that the nonrelativistic limits of the scalar density and the vector charge coincide), the known value of the pion-nucleon sigma term suggests that each quark carries a scalar density of 3$-$4, which is larger than the bare value 1.
This discrepancy cannot be fully explained by the existence of the disconnected quark-loop contribution, which is suggested by recent lattice QCD results to contribute less than 40\% to the sigma term \cite{sigmatermlattice}.
The above value of the pion-nucleon sigma term, although depending on the renormalization point, suggests that the dynamical scalar density of the quark $\langle N | \bar q q | N \rangle$ is enhanced from the bare one.
The physical meaning of the scalar density itself is ``how much and wide one can find particles and antiparticles''.
One can thus conjecture that strengthening the interquark potential, especially quark confinement, can suppress the quark scalar density.
One of the important objects of our study is to clarify and confirm this statement as an important physical signification of the quark scalar density.
In the Schwinger-Dyson formalism, it is actually possible to suggest this qualitative feature of this quantity.

Another important quantity is the quark axial vector charge of the nucleon, which gives the spin contribution of the quark to the nucleon.
The quark axial charge is given by the leading moment of the helicity distribution $g_1 (x)$ of the quark carrying the momentum fraction $x$ of the total momentum of the longitudinally polarized nucleon in the collinear factorization, and can be studied in high energy deep inelastic lepton-nucleon scattering experiments \cite{spinreview}.
In the nonrelativistic constituent quark model, one considers three massive quarks in the nucleon, thus reducing the axial charge of the quark to its spin.
In the proton, the spin fraction of the $u$ quark is then $\Delta u = \frac{4}{3}$, and for the $d$ quark we have $\Delta d = -\frac{1}{3}$.
Due to the success of the quark model, it was long thought that the quark spin carries the whole spin of the nucleon, but the European Muon Collaboration reported that the quark contribution is much smaller than 1 \cite{emc}.
This fact has surprised many physicists, and many theoretical studies trying to explain this ``proton spin crisis" have been done \cite{spinreview,spinmodel,spinanomaly,cheng-li_prl,strangespinellis} with no definitive consensus between them.
Currently the experimental studies continue \cite{helicityexp}, and the recent experimental data of the sum of the quark spin contribution to the nucleon spin is given by \cite{recentcompass}
\begin{eqnarray}
\Delta \Sigma &=& 0.32 \pm 0.03 \pm 0.03 
\, .
\label{eq:isoscalaraxialexp}
\end{eqnarray}
On the other hand, the isovector axial charge measured in neutron beta decay experiments is \cite{ucn}
\begin{eqnarray}
g_A &=& -1.27590 \pm 0.00239\, ^{+0.00331}_{-0.00377}  
\, .
\label{eq:isovectoraxialexp}
\end{eqnarray}
Many results of lattice QCD analyses are also available, which agree qualitatively with the above experimental data \cite{sigmatermandspinlattice,spinlattice}.
Despite the theoretical uncertainty, these two results show a suppression compared with the constituent quark model prediction.
This fact has to be explained from the point of view of the gauge-invariant angular momentum decomposition of the nucleon spin, which has recently received much theoretical development \cite{angularmomentum}.

The final charge of interest is the quark pseudoscalar density.
Despite its nonrelativistic suppression, the effect of the pseudoscalar density is phenomenologically known to be sizable due to the large value of the matrix element itself \cite{hai-yang_cheng,cheng-li_prl,edm2,herczeg2,isovectorpseudoscalardensity}.
This matrix element is also an important input in the search for new physics beyond the standard model \cite{darkmatter,edm2,herczeg2,hai-yang_cheng}.

When comparing the enhancement of the quark scalar/pseudoscalar densities and the suppression of the quark axial vector charge extracted from the experimental data or from lattice QCD simulations with the constituent quark model prediction, two sources of suppression can na\"{i}vely be inferred.
The first source is the dressing of the bare quark charges by gluons, and the second possibility is the hadronic bound state effect.
The second case was often discussed in the context of the quark model, but the first case was not discussed previously, and should be treated nonperturbatively to extract the physics.

We use the Schwinger-Dyson (SD) formalism as a powerful nonperturbative way to investigate the dynamics of quantum field theory and in particular low energy QCD;
many quantities have been investigated using this technique, such as the dynamical quark mass, the meson masses, etc \cite{higashijima,miransky,kaoki,robertsreview,alkoferreview,pinchtech,pi-k,emformfactor,quarkpropagator,unquenching,quark-gluon_vertex,tensorsde}.
In a previous work, we calculated the effect of the gluon dressing on the single quark tensor charge and showed that the gluon dressing suppresses the tensor charge of the bare quark, due to the superposition of states with spin-flipped quarks.
The effect in question, the vertex gluon dressing, is also well within the applicability of the SD formalism.
In this paper, we will therefore try to clarify the effect of the gluon vertex dressing and analyze the source of the deviation of the quark charges.

This paper is organized as follows.
In Section \ref{sec:setup}, we give the formulation of the SD formalism, the renormalization group (RG)-improved running couplings used in this work, and a brief explanation of the derivation of the dynamical quark mass.
In Section \ref{sec:scalar}, we formulate and solve the SD equation (SDE) for the quark scalar density and analyze the results.
A comparison between the analyses using the confining and nonconfining potentials is done to see the sensitivity of the quark scalar density on the confinement.
In Section \ref{sec:axial}, we formulate and solve the SDE for the quark axial charge.
The result of the quark axial charge SDE will be discussed by separating the gluon dressing effect and the quark-loop contribution related to the axial anomaly.
In Section \ref{sec:pseudoscalar}, we give the formula of the quark pseudoscalar density SDE and discuss its result .
The final section is devoted to the summary.

\section{\label{sec:setup}Basic Formalism}

In this section, we present the details of the SD formalism of Landau gauge QCD and the quark propagator used in this paper.
We consider the rainbow-ladder approximation where the nonperturbative effect of the gluon is included by improving the momentum dependence of the quark-gluon vertex \cite{quark-gluon_vertex} and the gluon dressing function by the one-loop level renormalization group.
This gives the replacement:
\begin{equation}
\frac{g_s^2}{4\pi} Z_g (q^2) \gamma^\mu \times \Gamma^\nu (q , k ) 
\rightarrow
\alpha_s (q^2) \gamma^\mu \times \gamma^\nu
\, ,
\label{eq:rlapprox}
\end{equation}
where $Z_g (q^2)$ is the gluon dressing function, and $\Gamma^\nu (q,k) $ is the dressed quark-gluon vertex.
In this work, we use the Landau gauge which minimizes the unphysical momentum fluctuation of the gluons in the Euclidean space-time.
We choose the RG-improved strong coupling with infrared (IR) regularization \`{a} la Higashijima (one-loop level, $N_f =3$) \cite{higashijima}.
The QCD scale parameter is fixed at $\Lambda_{\rm QCD}$ = 900 MeV 
(the ordinary QCD scale parameter is around $\Lambda_{\rm QCD} \simeq 200 - 300$ MeV. In this paper, the large scale parameter is taken to reproduce the chiral quantities).

\begin{figure}[htb]
\includegraphics[width=6.1cm,angle=-90]{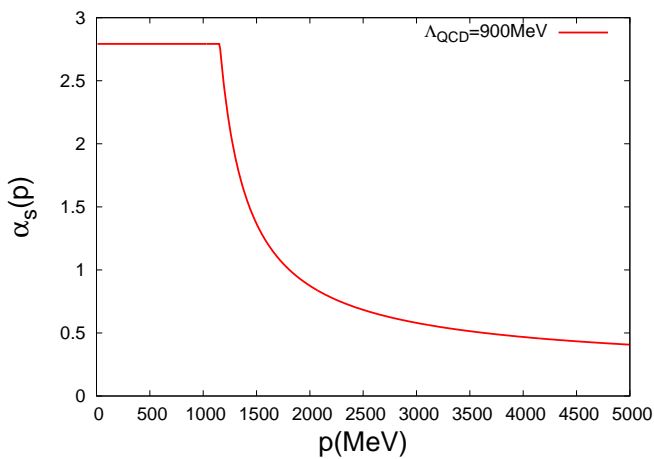}
\caption{\label{fig:strong_coupling}
The running strong coupling used in this work.
We use the running coupling with the simple infrared regularization with $\Lambda_{\rm QCD}=900$ MeV.
}
\end{figure}

The running strong coupling with the simple IR regularization is defined by \cite{higashijima}
\begin{eqnarray}
\alpha_s (p^2) =
\left\{
\begin{array}{ll}
\frac{8\pi}{\beta_0} & (p<p_{\rm IR}) \cr
\frac{4\pi}{\beta_0} \frac{1}{\ln (p^2 / \Lambda_{\rm QCD}^2)} & (p\geq p_{\rm IR}) \cr
\end{array}
\right.
\, ,
\label{eq:simpleircut}
\end{eqnarray}
where $\beta_0 = \frac{11N_c -2N_f}{3}$.
Here we take $N_c =N_f =3 $ and $p_{\rm IR}$ satisfying $\ln (p_{\rm IR}^2 / \Lambda_{\rm QCD}^2)= \frac{1}{2}$.
As can be seen in Fig. \ref{fig:strong_coupling}, this running coupling has one cusp in the infrared region.
This IR regularization was introduced to avoid the divergent Landau pole at $p=\Lambda_{\rm QCD}$.
The shape of the running coupling is plotted in Fig. \ref{fig:strong_coupling}.

\begin{figure}[htb]
\begin{center}
\includegraphics[width=6.1cm,angle=-90]{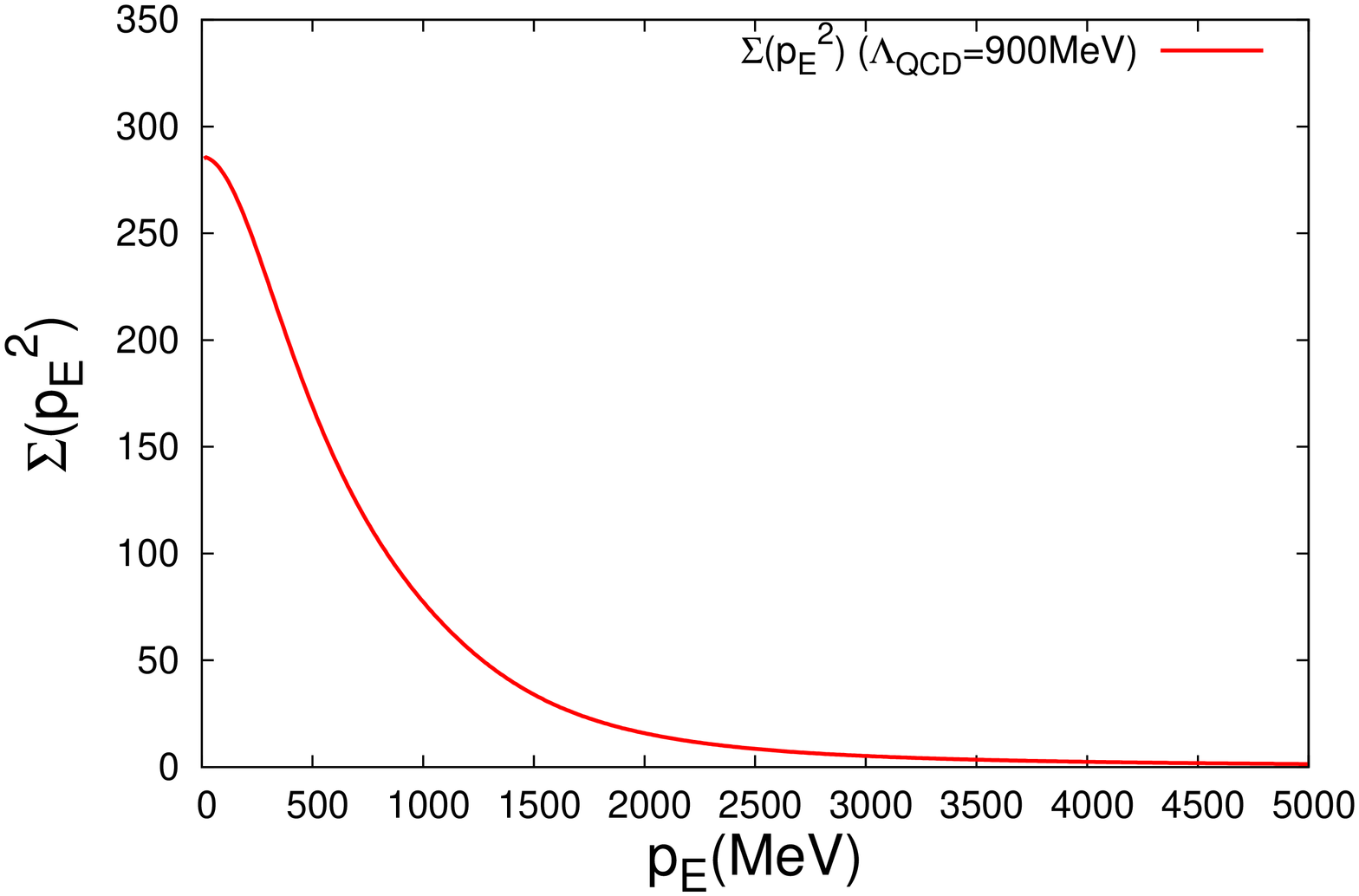}
\caption{\label{fig:self-energy}
The quark self-energy $\Sigma (p_E^2)$ solved with the Schwinger-Dyson equation in the chiral limit.
}
\end{center}
\begin{center}
\includegraphics[width=6.1cm,angle=-90]{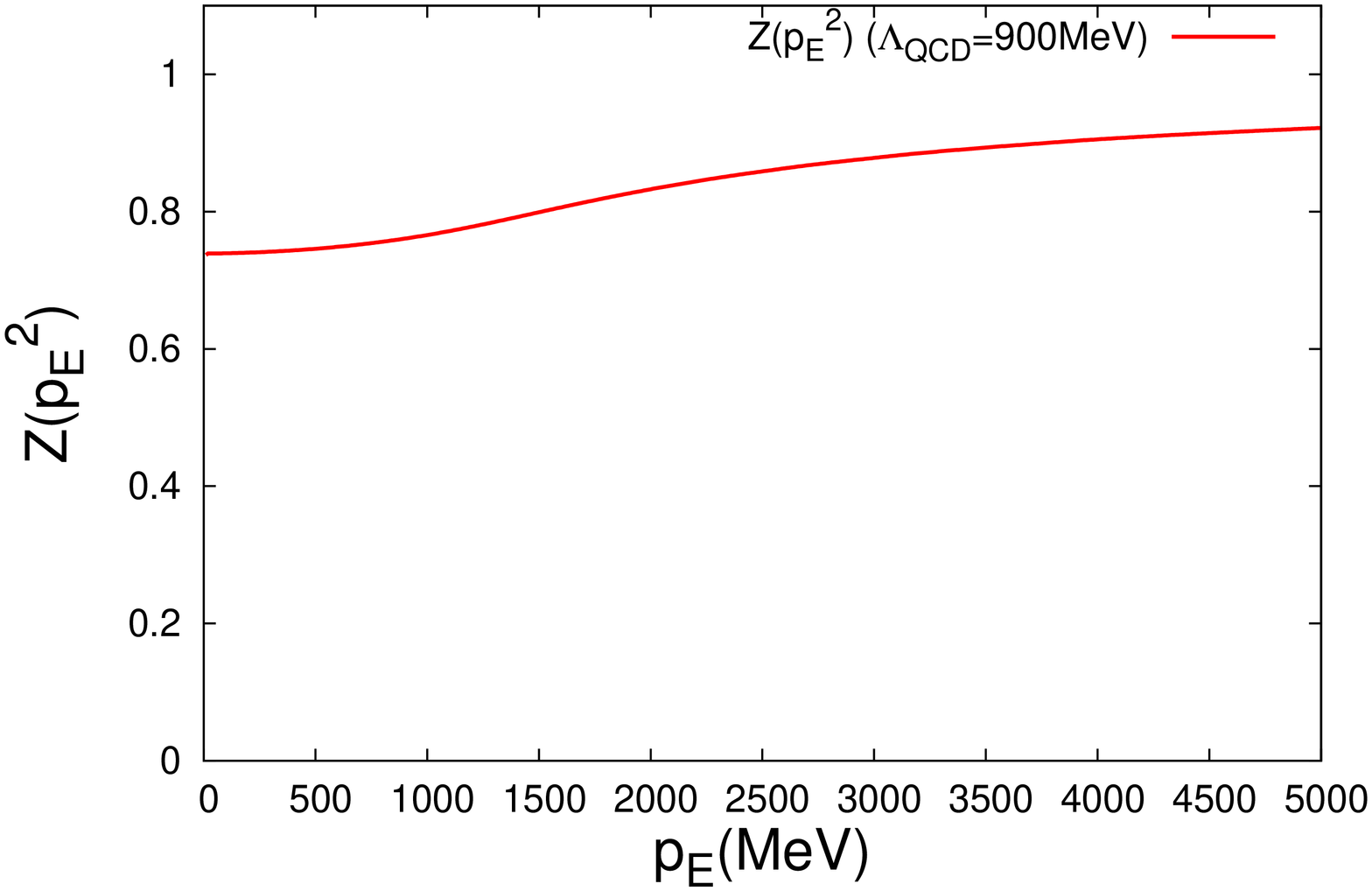}
\caption{\label{fig:wavfctnrenormalization}
The quark wave function renormalization $Z (p_E^2)$ solved with the Schwinger-Dyson equation in the chiral limit.
}
\end{center}
\end{figure}

We now solve the quark propagator SDE in Landau gauge QCD.
In this paper, we consider the SDE with the effect of the dressed gluon propagator and the dressed quark-gluon vertex included in the RG-improved strong coupling [see Eq. (\ref{eq:rlapprox})].
The SDE is a system of two integral equations:
\begin{eqnarray}
\frac{\Sigma (p^2)}{Z(p^2)} 
&=&
m_q (\Lambda)
-\frac{3i \,C_2 (N_c)}{4\pi^3} \int \hspace{-.3em} d^4k
\frac{\alpha_s [(p-k)^2]}{(p-k)^2}
\nonumber\\
&&\hspace{10em} \times
\frac{Z(k^2) \Sigma (k^2)}{k^2 -\Sigma^2 (k^2)}
.
\label{eq:SD_eq4}
\\
\frac{1}{Z (p^2)}
&=&
1+i \frac{C_2 (N_c) }{8\pi^3 p^2} 
\int \hspace{-.3em} d^4k
\frac{ \alpha_s [(p-k)^2] }{k^2 -\Sigma^2 (k^2)}
Z(k^2)
\nonumber\\
&&\hspace{3em} \times
\left[
2-\frac{p^2+k^2}{(p-k)^2} -\frac{(p^2-k^2)^2}{(p-k)^4}
\right]
.
\label{eq:wavfctnsde}
\end{eqnarray}
where $Z(k^2)$ and $\Sigma (k^2)$ are the wave function renormalization and the self-energy of the quark, respectively, and $C_2 (N_c) \equiv \sum_a^{N_c^2 -1} T_a T_a = \frac{N_c^2 - 1}{2N_c}$ is the Casimir operator of the $SU(N_c)$ group.
Here the current quark mass $m_q (\Lambda)$ is to be understood as the bare quark mass defined at the scale of the integral cutoff.
The bare quark mass used in the above equation expressed in terms of the current quark mass at the renormalization point $\mu$ is therefore given as
\begin{equation}
m_q (\Lambda )
=
\left( \frac{\alpha_s (\Lambda^2 )}{\alpha_s (\mu^2 ) } \right)^{\frac{3C_2 (N_c)}{\beta_0}}
m_q (\mu )
\, ,
\label{eq:renormalized_current_quark_mass}
\end{equation}
where $\Lambda$ is the integral cutoff and $\frac{3C_2 (N_c)}{ \beta_0}=\frac{4}{9}$. 
The quark wave function renormalization and the quark self-energy in the chiral limit are plotted in Figs. \ref{fig:self-energy} and \ref{fig:wavfctnrenormalization}, respectively.
We see that the self-energy is generated dynamically even in the chiral limit.
There the value of the self-energy at $p=0$ is given by $\Sigma (p_E =0) = 285$ MeV.
This dynamically generated quark mass can be seen as the constituent quark mass in the quark model.

The quark self-energy can be related to the chiral condensate with
\begin{equation}
\langle \bar q q \rangle_\Lambda 
=
-
\frac{N_c }{ 2\pi^2} \int_0^{\Lambda} 
\hspace{-0.5em}
k_E^3 dk_E
\frac{Z (k_E^2) \Sigma(k_E^2)}{k_E^2 + \Sigma^2 (k_E^2)}
\, ,
\label{eq:chiral_condensate}
\end{equation}
where $\Lambda$ is the ultraviolet cutoff (not to be confused with $\Lambda_{\rm QCD}$).
In our numerical calculation, the cutoff was taken as $\Lambda =$ 10 GeV.
To  obtain the chiral condensate renormalized at $\mu =$ 2 GeV, we use the formula
\begin{equation}
\langle \bar q q \rangle_\mu
=
\left( \frac{\alpha_s (\Lambda^2 )}{\alpha_s (\mu^2 ) } \right)^{\frac{3C_2 (N_c)}{ \beta_0}}
\langle \bar q q \rangle_\Lambda
\, .
\label{eq:renormalized_chiral_condensate}
\end{equation}
The above renormalized chiral condensate is stable in the variation of the cutoff scale $\Lambda$ [numerically, we have verified that the variation is small, of $O(10^{-3})$]. 
The numerical value is $\langle \bar q q \rangle_{\mu = 2\, {\rm GeV}} \approx -(237\, {\rm MeV})^3$ for the chiral limit.

From the quark self-energy, it is also possible to give the pion decay constant $f_\pi$ with the Pagels-Stokar approximation \cite{pagels}:
\begin{eqnarray}
f_\pi^2 
&=&
\frac{N_c }{2\pi^2} \int_0^\infty \hspace{-0.5em}k_E^3 dk_E \, \frac{\Sigma (k_E^2 ) Z(k_E^2)}{\left[ k_E^2 +\Sigma^2 (k_E^2 ) \right]^2}
\nonumber\\
&&\hspace{6em} \times
\left[ \Sigma (k_E^2 ) - \frac{k_E}{4} \frac{d}{dk_E} \Sigma (k_E^2 ) \right]
. \ \ \ 
\label{eq:pagels-stokar}
\end{eqnarray}
The pion decay constant is an observable, so its renormalization is not required.
In the chiral limit, we obtain $f_\pi = 66$ MeV.
We note that the experimental value of the pion decay constant is $f_\pi = 93$ MeV.

\section{\label{sec:scalar}Quark scalar density}

\subsection{\label{sec:scalarsdeq}Quark scalar density: \\the Schwinger-Dyson equation}

\begin{figure}[htb]
\includegraphics[width=8.2cm]{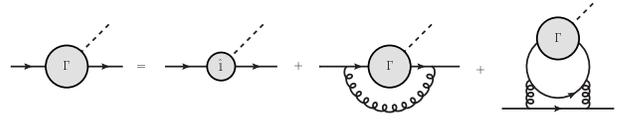}
\caption{\label{fig:scalar_SD_eq}
The Schwinger-Dyson equation for the quark scalar density expressed diagrammatically.
The last two-loop diagram is the isoscalar unquenching effect, which is absent for the isovector quark scalar density.
}
\end{figure}

Let us consider the SDE of the quark scalar density depicted diagrammatically in Fig. \ref{fig:scalar_SD_eq}.
The SDE for the quenched (isovector) quark scalar density is given by
\begin{eqnarray}
\Gamma (p)
&=&
1
\nonumber\\
&&
+
i C_2 (N_c) \int \frac{d^4k}{4\pi^3}
\alpha_s \left[(p-k)^2 \right] D_{\rho \lambda} (p-k)
\nonumber\\
&& \times
Z^2(k^2)
\gamma^\rho \frac{k\hspace{-.45em}/\, + \Sigma (k^2)}{k^2 -\Sigma^2 (k^2)}
\Gamma (k)
\frac{k\hspace{-.45em}/\, + \Sigma (k^2)}{k^2 -\Sigma^2 (k^2)} \gamma^\lambda 
,
\nonumber\\
\label{eq:scalarsde}
\end{eqnarray}
where $ D_{\rho \lambda} (q) \equiv \frac{-1}{q^2} \left( g_{\rho \lambda} - \frac{q_\rho q_\lambda }{q^2} \right) $ is the gluon propagator in the Landau gauge (the color index was factorized), and $\Gamma$ is the dynamical scalar density in the zero limit of the momentum transfer.
As for the quark propagator SDE, we consider the rainbow-ladder approximation [see Eq. (\ref{eq:rlapprox})] in which the effect of the dressed gluon propagator and the dressed quark-gluon vertex are included in the RG-improved strong coupling given in the previous section.
In this section, we consider the chiral limit $m_q=0$.

\begin{figure}[htb]
\includegraphics[width=12cm,angle=-90]{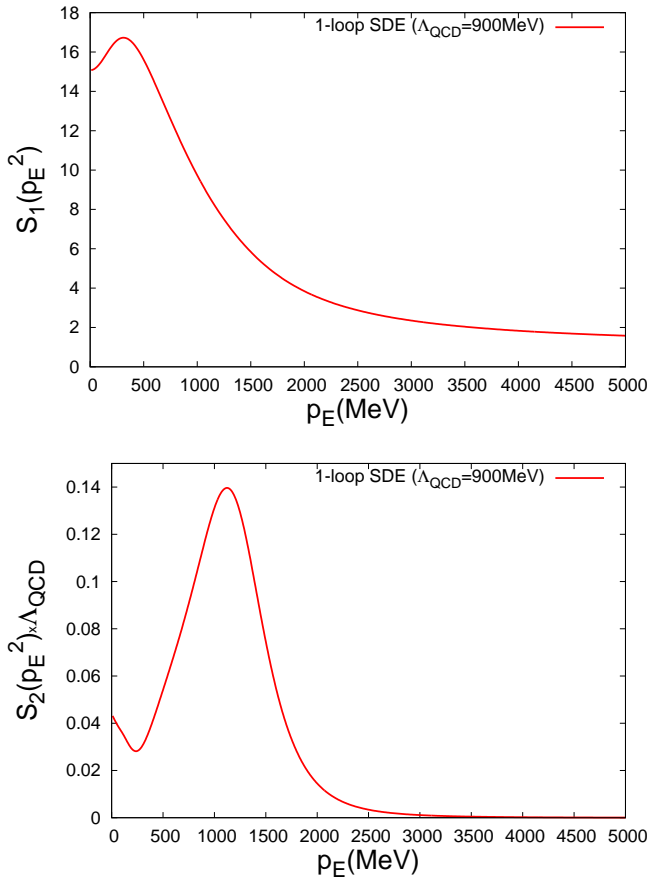}
\caption{\label{fig:scalar_charge_12}
The $S_1$ and $S_2$ functions (not renormalized) solved with the Schwinger-Dyson equation for the quark scalar density with the integral cutoff $\Lambda = 10$ GeV.
}
\end{figure}

In Eq. (\ref{eq:scalarsde}), there are two relevant Lorentz structures: $\hat 1$ and $p\hspace{-.45em}/\, $.
The dynamical scalar density is thus written as 
\begin{equation}
\Gamma (p) 
\equiv 
S_1 (p^2) 
+S_2 (p^2)  p\hspace{-.45em}/\,
\, .
\label{eq:scalarstructure}
\end{equation}
The SDE (\ref{eq:scalarsde}) can thus be rewritten as a set of integral equations with the $S_1 (p^2)$ and $S_2 (p^2)$ functions.
The zero momentum point of the $S_1$ function indicates the ratio between the scalar density of the dressed and bare quarks. 
After some algebra, we find the following set of integral equations:
\begin{widetext}
\begin{eqnarray}
S_1 (p_E^2)
&=&
1
+\frac{3 C_2 (N_c)}{\pi^2} \int_0^\infty \hspace{-0.5em} k_E^3 dk_E \int_0^\pi \hspace{-0.3em} \sin^2 \theta d\theta
\frac{\alpha_s (p_E^2+k_E^2 -2 p_E k_E \cos \theta ) \, Z^2 (k_E^2) }{(p_E^2+k_E^2 -2 p_E k_E \cos \theta ) \left[ k_E^2 +\Sigma^2 (k_E^2) \right]^2} 
\nonumber\\
&& \hspace{17em} \times 
\Biggl\{ \, 
\left[ k_E^2 - \Sigma^2 (k_E^2) \right] S_1 (k_E^2) +2 k_E^2 \Sigma (k_E^2) S_2 (k_E^2)
\ \Biggr\}
\ ,
\label{eq:sdes1}
\\
S_2 (p_E^2)
&=&
\frac{ C_2 (N_c) }{\pi^2 p_E^2 } \int_0^\infty \hspace{-0.5em} k_E^3 dk_E \int_0^\pi \hspace{-0.3em} \sin^2 \theta d\theta
\frac{\alpha_s (p_E^2+k_E^2 -2 p_E k_E \cos \theta ) \, Z^2 (k_E^2) }{ \left[ k_E^2 +\Sigma^2 (k_E^2) \right]^2} 
\nonumber\\
&& \hspace{9em} \times 
\left[ 
\frac{(p_E^2-k_E^2)^2}{(p_E^2+k_E^2 -2 p_E k_E \cos \theta )^2}
+\frac{p_E^2+k_E^2}{(p_E^2+k_E^2 -2 p_E k_E \cos \theta )}
-2
\right] 
\nonumber\\
&& \hspace{18em} \times 
\Biggl\{ \, 
\Sigma (k_E^2) S_1 (k_E^2)- \frac{1}{2} \left[ k_E^2 - \Sigma^2 (k_E^2) \right] S_2 (k_E^2) 
\ \Biggr\}
\ ,
\label{eq:sdes2}
\end{eqnarray}
where we have used the Wick rotated momenta.
For the derivation of the above integral equations, see Appendix \ref{sec:scalar_derivation_1-loop}.
The result of the SDE for the quark scalar density is plotted in Fig. \ref{fig:scalar_charge_12}.

To discuss the unquenched (isoscalar) quark scalar density, we must extend the SDE of Eq. (\ref{eq:scalarsde}).
In our discussion, we have considered the quark-loop contribution as the leading unquenching effect (see the two-loop diagram in Fig. \ref{fig:scalar_SD_eq}).
This corresponds to partially including the effect of the disconnected quark-loop contribution in the language of lattice QCD.
The SDE of the unquenched (isoscalar) scalar density is given by
\begin{eqnarray}
S_1 (p_E^2) 
&=&
\frac{3C_2(N_c)}{ \pi^2} \int_0^\Lambda \hspace{-0.5em} dk_E \frac{\alpha_s^2 (k_E^2)}{k_E} \int_0^\pi \hspace{-0.5em} \sin^2 \theta d \theta
\frac{Z[(p_E -k_E)^2] \Sigma [(p_E -k_E)^2] }{(p_E -k_E )^2 +\Sigma^2[(p_E -k_E )^2]} f \bigl[S_1,S_2 ; k_E^2 \bigr]
+[\mbox{RHS of Eq. (\ref{eq:sdes1})}]
, \ \ 
\label{eq:sdes1_ql}
\end{eqnarray}
\begin{eqnarray}
S_2 (p_E^2) 
&=&
\frac{C_2(N_c)}{2 \pi^2 p_E^2} \int_0^\Lambda \hspace{-0.5em} dk_E \frac{\alpha_s^2 (k_E^2)}{ k_E^3} \int_0^\pi \hspace{-0.5em} \sin^2 \theta d \theta
\frac{Z[(p_E -k_E)^2]}{(p_E -k_E )^2 +\Sigma^2 [(p_E -k_E )^2]} f \bigl[S_1,S_2 ; k_E^2 \bigr]
\nonumber\\
&& \hspace{6.8em} \times 
\Biggl[ 2k_E^4 -p_E^2 k_E^2 -p_E^4 + (2p_E^2 -k_E^2 ) (p_E -k_E)^2 - (p_E -k_E)^4 \Biggr]
+[\mbox{RHS of Eq. (\ref{eq:sdes2})}]
,
\label{eq:sdes2_ql}
\end{eqnarray}
where the function $f \bigl[S_1,S_2 ; k_E^2 \bigr]$ is defined as
\begin{eqnarray}
f \bigl[S_1,S_2 ; k_E^2 \bigr]
&\equiv &
\frac{4 N_f }{\pi^2 } \int_0^\Lambda \hspace{-0.5em} l_E^3 dl_E \int_0^\pi \hspace{-0.5em} \sin^2 \theta d\theta
\frac{Z[(k_E -l_E)^2] \, Z^2 (l_E^2)}{\Bigl[ (k_E -l_E )^2 +\Sigma^2 [(k_E -l_E )^2] \Bigr] \Bigl[ l_E^2 +\Sigma^2 (l_E^2 ) \Bigr]^2 } 
\nonumber\\
&& \hspace{3em}
\times \Biggl\{
\ S_1(l_E^2) 
\Biggl[ 
\Bigl[ l_E^2 - \Sigma^2 (l_E^2) \Bigr]
\Sigma[(l_E-k_E )^2]
\nonumber\\
&& \hspace{8.5em}
+\frac{\Sigma (l_E^2) }{3k_E^2 } \Bigl[ 2k_E^4 -l_E^2 k_E^2 -l_E^4 + (2l_E^2 -k_E^2 )(l_E-k_E)^2 -(l_E -k_E)^4 \Bigr]
\Biggr]
\nonumber\\
&& \hspace{4.5em}
+S_2(l_E^2) 
\Biggl[ 
\, 2 l_E^2 \Sigma (l_E^2) \Sigma[(l_E-k_E )^2]
\nonumber\\
&& \hspace{8.5em}
-\frac{l_E^2 - \Sigma^2 (l_E^2)}{6k_E^2 } 
\Bigl[ 2k_E^4 -l_E^2 k_E^2 -l_E^4 + (2l_E^2 -k_E^2 )(l_E-k_E)^2 -(l_E -k_E)^4 \Bigr]
\Biggr]
\, \Biggr\}
, \ \ \ \ \ 
\label{eq:scalarinnerloopfunction}
\end{eqnarray}
\end{widetext}
where $(p_E -k_E)^2 \equiv p_E^2+k_E^2 -2 p_E k_E \cos \theta $.
For the derivation of the above integral equations, see Appendix \ref{sec:scalar_derivation_2-loop}.
The unquenched (isoscalar) quark scalar density SDE (\ref{eq:sdes1_ql}) and (\ref{eq:sdes2_ql}) do not converge in our setup.
We will analyze this fact in the next subsection.

\subsection{\label{sec:scalaranalysis}Quark scalar density: Analysis}

By looking at the solution of the quenched (isovector) quark scalar density SDE in Fig. \ref{fig:scalar_charge_12}, we see that the scalar density of the quark is significantly enhanced when the scalar vertex is dressed by the gluons.
The $S_1$ and $S_2$ functions obtained after solving Eqs. (\ref{eq:sdes1}) and (\ref{eq:sdes2}) are dependent on the cutoff $\Lambda$, and we need to renormalize the tensor charge at some fixed scale.
At the scale $\mu$, the renormalized quark scalar density is given by 
\begin{equation}
S_1 (0)_\mu 
=
\left( \frac{\alpha_s (\Lambda^2 )}{\alpha_s (\mu^2 ) } \right)^{-\frac{3C_2 (N_c)}{\beta_0}}
S_1 (0)_\Lambda
\, ,
\label{eq:scalar_density_renormalization}
\end{equation}
where $S_1 (0)_\Lambda$ is the scalar density given as the solution of the cutoff ($\Lambda$) dependent SDE.
The exponent is $-\frac{3C_2 (N_c)}{ \beta_0}=-\frac{4}{9}$ for $N_c = 3$ and $N_f =3$, the same as for the chiral condensate (note that $m_q \langle N|\bar q q|N\rangle$ is renormalization independent).

From the above formula, we obtain the renormalized quenched (isovector) quark scalar density at $\mu =2$ GeV
\begin{eqnarray}
S_1 (0)_{\mu =2\, {\rm GeV}} &=& 9.2  \, .
\label{eq:scalarresult}
\end{eqnarray}
We see that the renormalized $S_1$(0) is larger than 1.
This fact shows that the scalar density of the dressed quark is enhanced from the bare quark contribution by the gluon dressing of the vertex.

\begin{figure}[htb]
\includegraphics[width=6.1cm,angle=-90]{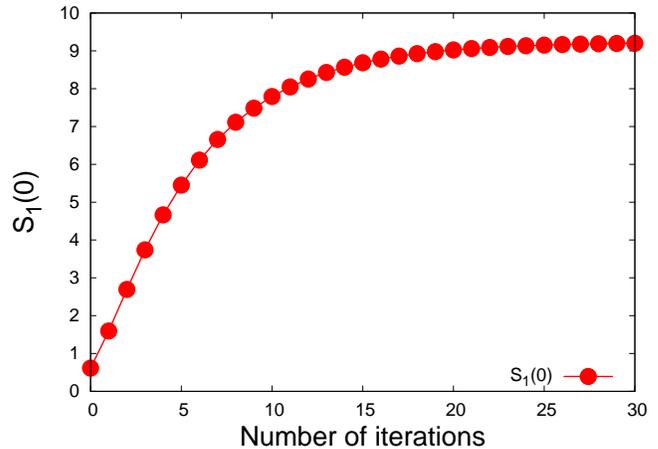}
\caption{\label{fig:scalar_s1(0)}
The convergence of the isovector quark scalar density $S_1(0)$ vs the number of iterations of the Schwinger-Dyson equation with the initial conditions $S_1(p_E^2) = 1$ and $S_2(p_E^2) = 0$.
The scalar density was renormalized at $\mu = 2$ GeV.
}
\end{figure}

\begin{figure}[htb]
\includegraphics[width=6cm]{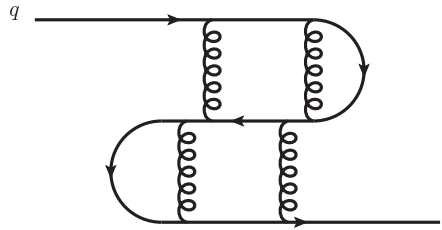}
\caption{\label{fig:zgraph}
The Z graph of the quark propagation with time in the horizontal direction.
This describes quark pair creation and annihilation in the intermediate state.
The propagation of the virtual quark pairs are restricted by the interquark potential due to the gluon exchange.
}
\end{figure}

Let us try to understand the enhancement of the quark scalar density with the gluon vertex dressing.
The structure of the quark scalar density SDE (see Fig. \ref{fig:scalar_SD_eq}) shows that the successive iteration of the substitution of the left-hand side of the SDE into its right-hand side yields a sort of perturbative expansion in which the number of iterations corresponds to the order of perturbation. 
The quark scalar density $S_1 (0)$ obtained after each iteration is shown in Fig. \ref{fig:scalar_s1(0)}.
From Fig. \ref{fig:scalar_s1(0)}, we can see that the isovector quark scalar density converges by increasing monotonically.
This means that the gluon emission/absorption by the quark increases the quark scalar density.
This fact can be easily understood since the scalar density of the quark is the sum of the probability of finding a quark in the intermediate state in the whole space-time.
The gluon emission and absorption extend the phase space of the quark so that the configuration of the quark propagation becomes larger than that for the noninteracting single quark.
It is important to note that the extension of the possible path of the quark also extends in the time direction.
This comprises the case of the particle/antiparticle pair creation as shown in the Z graph of Fig. \ref{fig:zgraph}.
The particle and antiparticle propagations contribute with the same sign to the scalar density.
This fact is in contrast to the vector current.

From the phenomenology and lattice QCD analyses, the pion-nucleon sigma term is given by $m_q\langle N | \bar q q | N \rangle \sim 45$ MeV.
The quark scalar density in the nucleon is therefore $\langle N | \bar q q | N \rangle \sim 14$ at the renormalization scale $\mu =2$ GeV, where we have used $m_q \sim 3.5$ MeV \cite{pdg}.
The recent lattice QCD results suggest that the quark-loop contribution to the sigma term represents the 20-30\% of the total magnitude.
From this, the quenched part of the sigma term is $\langle N | \bar q q | N \rangle_{\rm disc} \sim 10$.
If we assume that the scalar density of the single quark is given by a third of $\langle N | \bar q q | N \rangle_{\rm disc}$ (this assumption relies on the fact that the nonrelativistic limits of the vector charge and the scalar density coincide), we obtain $S_1 (0)_{\mu =2 \, {\rm GeV}} \sim 3$.
This value is well below our result from Eq. (\ref{eq:scalarresult}).
How are we to understand this result?
In the previous paragraph, we have seen that the scalar density is the sum of the probability of finding a quark in the intermediate state in the whole space-time and that the quark scalar density becomes large if the configuration of the quark propagation in the intermediate state is large.
This suggests that the quark scalar density becomes smaller when the quark is affected by a stronger attractive potential, especially the confining potential.
The nucleon matrix element $\langle N | \bar q q | N \rangle_{\rm disc} \sim 10$ gives the scalar density of the quarks confined in the nucleon, so it is natural to find a smaller value than what we have considered since we have not considered the effect of the quark confinement.

To partially see the effect of the confined quark on the scalar density, we can make use of the phenomenological running coupling of Richardson \cite{richardson}.
The Richardson Ansatz assumes the running coupling
\begin{equation}
\alpha_s (p^2)
=
\frac{4\pi}{\beta_0} \frac{1}{\ln (1+p^2 / \Lambda_{\rm QCD}^2)} 
\, ,
\end{equation}
instead of Eq. (\ref{eq:simpleircut}), and it can generate a linearly confining potential.
Due to the divergent pole at $p=0$, it is not possible to calculate the quark propagator SDE with the Richardson Ansatz in our setup, so we must use some regularization.
Using the Higashijima-Miransky approximation
\begin{equation}
\alpha_s [(p_E -k_E )^2]
\approx
\alpha_s [{\rm max} (p_E^2 , k_E^2 ) ]
\, ,
\end{equation}
it is possible to regularize the divergence along the line $p_E=k_E$ in the phase space.
It is thus possible to analyze qualitatively the effect of quark confinement by comparing the quark scalar density obtained with and without the Richardson Ansatz in the Higashijima-Miransky approximation.
By using the above Richardson Ansatz in our SD formalism, it is possible to partially include the effect of quark confinement on the single quark.
This corresponds to restricting the path of the quark propagation in the Z graph (see Fig. \ref{fig:zgraph}).

\begin{figure}[htb]
\includegraphics[width=12cm,angle=-90]{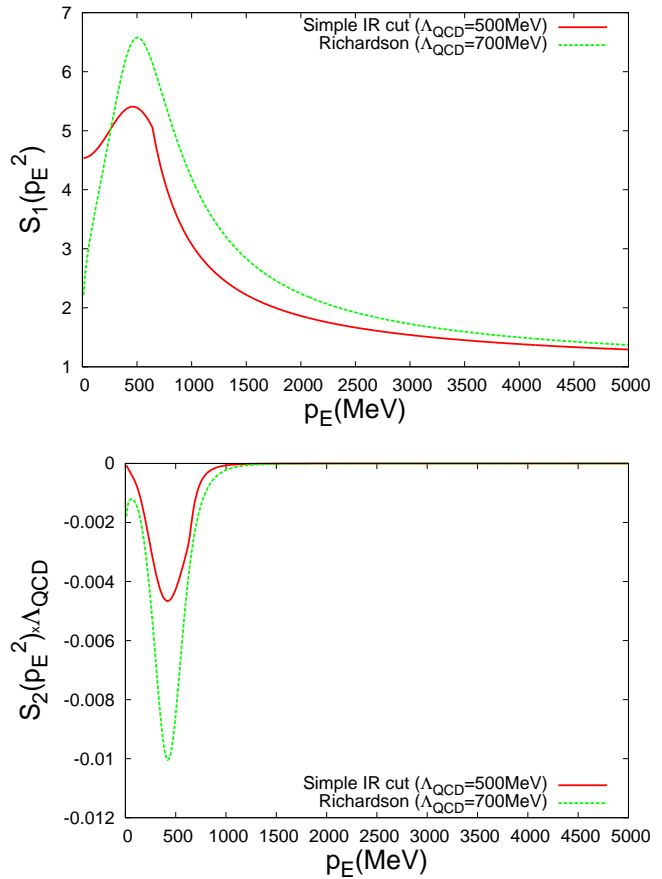}
\caption{\label{fig:scalar_charge_hm_12}
Comparison of the $S_1$ and $S_2$, functions (not renormalized) solved with the quark scalar density SDE with the integral cutoff $\Lambda = 10$ GeV in the Higashijima-Miransky approximation with and without the Richardson Ansatz.
}
\end{figure}

In the evaluation of the case without quark confinement, we have used the simple IR cutoff introduced in Section \ref{sec:setup}, with the QCD scale parameter $\Lambda_{\rm QCD} = 500$ MeV.
With this input, the quark propagator SDE (\ref{eq:SD_eq4}) and (\ref{eq:wavfctnsde}) exhibits $\langle \bar q q \rangle_{\mu = 2\, {\rm GeV}} = -(242 \, {\rm MeV})^3$ for the chiral condensate and $f_\pi = 91$ MeV for the pion decay constant.
By calculating the quark scalar density SDE (\ref{eq:sdes1}) and (\ref{eq:sdes2}), we obtain
\begin{equation}
S_1 (0)_{\mu =2\, {\rm GeV}}
=
3.2
\, .
\end{equation}
Using the Richardson Ansatz with $\Lambda_{\rm QCD} = 700$ MeV (in this setup, we obtain $\langle \bar q q \rangle_{\mu = 2\, {\rm GeV}} = -(260 \, {\rm MeV})^3$ and $f_\pi = 93$ MeV), we have
\begin{equation}
S_1 (0)_{\mu =2\, {\rm GeV}}
=
1.5
\, .
\end{equation}
A comparison of the solutions of the quark scalar density SDE is shown in Fig. \ref{fig:scalar_charge_hm_12}.
We see that the quark scalar density is smaller for the case where the effect of quark confinement is included with the Richardson Ansatz.
This result strongly suggests that the quark scalar density is an observable that is sensitive to quark confinement.

We have seen that it is possible to explain the phenomenological value of the pion-nucleon sigma term.
Our approach is based on the quark model point of view.
This is not surprising since the quark propagator SDE generates a dynamical quark mass which associates the dressed quark with the massive constituent quark.

This property of the quark scalar density should also be notable for multi-hadron states, since we have more valence quarks and anitquarks.
In a scattering state, the quark scalar density of the multi-hadron state should be given by the sum of the scalar densities of the single states of each hadron, since the hadrons have small correlations with one other.
Similarly, for molecular systems the baryons and mesons are quasi-on-shell and well distant from one other, so we may expect that the scalar density is approximately given by the sum of the scalar densities of the hadronic components, and that it consequently drives the hadronic molecules to have a larger quark scalar density than the single multi-quark hadron system with the same quantum number.
Intuitively, this should be explained by the fact that the probability of finding an on-shell meson around molecular hadrons is much larger than that of finding an off-shell meson of the meson cloud of a baryon.
We can thus say that the quark scalar density is an observable that is sensitive to the compositeness of the hadrons, and we can provide a new approach to the problem of the structure of the hadrons, in addition to the classic approach \cite{weinberg}.

We also discuss the unquenching (isoscalar) quark-loop effect on the quark scalar density SDE.
In Eqs. (\ref{eq:sdes1_ql}) and (\ref{eq:sdes2_ql}), we have given the analytic formula of the quark-loop contribution, but the quark scalar density SDE does not converge in our setup.
Although many improvements such as the effect beyond the rainbow approximation \cite{unquenching}, the gluon sector SDE, the quark-gluon vertex SDE \cite{quarkpropagator,quark-gluon_vertex}, etc are required in our formalism, we will try to estimate the source of this failure.
In our SD formalism, the effect of the quark confinement was not included.
The quark of the unquenching loop can thus take a very large path so that the contribution to the scalar density becomes very large, upsetting the convergence.
It is also known that the quark-loop screens the interquark potential in the SD formalism \cite{quark-gluon_vertex,screening}, and can potentially cause bad infrared behavior of the inner loop integral of Eq. (\ref{eq:scalarinnerloopfunction}).
The many-body effect should also be very important in the evaluation of the quark-loop contribution, since the exclusion principle can be effective for a bound-state.

\begin{figure}[htb]
\includegraphics[width=6.1cm]{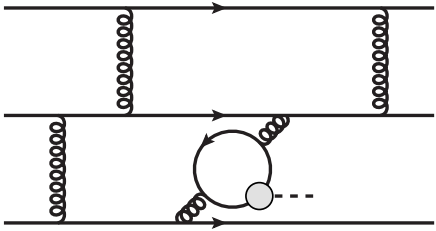}
\caption{\label{fig:exchangecurrent}
Schematic picture of the exchange current contribution from the unquenching quark-loop.
The gray blob represents the scalar operator insertion.
}
\end{figure}

To do a qualitative analysis of the unquenching quark-loop effect, a more detailed study is required.
The first possibility is to consider the contribution from the exchange current (see Fig. \ref{fig:exchangecurrent}), which is also expected to be sizable and not necessarily of the same sign as the quark-loop contributing to the scalar charge of the single quark.
It is also known that the pion-nucleon sigma term receives a sizable contribution from the pion cloud \cite{sigmatermmodel}.
The study of the pion (Nambu-Goldstone) mode as a higher order unquenching effect in the SD formalism \cite{unquenching} is also required.

\begin{figure}[htb]
\includegraphics[width=6.1cm,angle=-90]{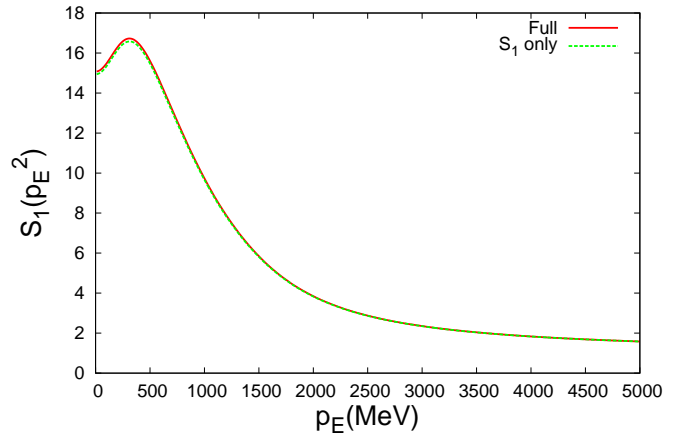}
\caption{\label{fig:scalar_s1_compare}
The $S_1$ function (not renormalized) obtained by solving the Schwinger-Dyson equation with $S_2$ function set to zero.
The $S_1$ function solved with the full contribution ($S_1$ and $S_2$) is also shown for comparison.
}
\end{figure}

In the formalism we have adopted, it is possible to change the input parameters and the self-energy function we have obtained in the intermediate steps, and this fact is an important advantage of the SD formalism.
We have tested the contribution of the $S_1$ and $S_2$ functions through a fictitious manipulation by setting $S_2 (p^2) =0$ when solving the SDE (\ref{eq:sdes1}) and (\ref{eq:sdes2}).
This approximation was tested previously in the analysis of the quark tensor charge, and the reduction of the SDE was successful within a few percent \cite{tensorsde}.
The result is plotted in Fig. \ref{fig:scalar_s1_compare}.
We see that the solutions of the SDE with and without the contribution from the $S_2$ function are very close.
The qualitative features are very similar.
This result suggests that the extra powers of the momenta $p$ (appearing in $p\hspace{-.45em}/\,$) work as a suppression factor. 
This shows that the leading contribution to the SDE of the quark tensor charge is given by the $S_1$ function, and that the omission of the $S_2$ function is a relatively good approximation.

We should also add that the dressed quark scalar density has a dependence on the scale parameter $\Lambda_{\rm QCD}$.
We show the coefficient $S_1 (0)$ renormalized at $\mu = 2$ GeV for several values of $\Lambda_{\rm QCD}$ in Table \ref{table:scalarlambdaqcddependence}.
We see that the scalar density increases as the scale parameter decreases.
This shows that the quark can propagate with a larger path when $\Lambda_{\rm QCD}$ is small.
This is quite natural, since a quark with small dynamical mass can move much more than a heavier one, and this situation is realized for smaller values of the QCD scale parameter.

\begin{table}
\caption{The quark scalar density obtained with several values of $\Lambda_{\rm QCD}$.
The renormalization point was fixed to $\mu = 2$ GeV.
}
\begin{ruledtabular}
\begin{tabular}{ccccc}
$\Lambda_{\rm QCD}$ & 200 MeV & 500 MeV & 900 MeV & 1 GeV \\
\hline
$S_1(0)_\mu$ & 14.6 & 11.3 & 8.61 & 6.36 \\
\end{tabular}
\end{ruledtabular}
\label{table:scalarlambdaqcddependence}
\end{table}

\section{\label{sec:axial}Quark axial charge}

\subsection{\label{sec:axialsdeq}Quark axial charge: Schwinger-Dyson equation}

\begin{figure}[htb]
\includegraphics[width=8.2cm]{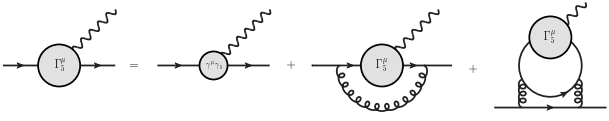}
\caption{\label{fig:axial_SD_eq} 
The Schwinger-Dyson equation for the quark axial charge expressed diagrammatically.
}
\end{figure}

We now consider the SDE of the quark axial charge. 
In our truncation scheme, we take into account the gluon dressing effect in the rainbow-ladder approximation as for the quark propagator SDE [see Eq. (\ref{eq:rlapprox})] and the leading quark-loop contribution as the unquenching effect.
In our approximation, we use the momentum of the gluon as the argument of the running coupling so that the chiral Ward identity is preserved \cite{kugomitchard}.
The quark axial charge SDE is depicted diagrammatically in Fig. \ref{fig:axial_SD_eq}.

We first treat the quark axial charge SDE without the quark-loop effect.
This corresponds to the SDE of the isovector quark axial charge.
It is given by
\begin{eqnarray}
\Gamma_5^\mu (p)
&=&
\gamma^\mu \gamma_5
\nonumber\\
&&
+
i C_2 (N_c) \int \frac{d^4k}{4\pi^3}
\alpha_s \left[(p-k)^2 \right] D_{\rho \lambda} (p-k)
\nonumber\\
&& \times
Z^2 (k^2)
\gamma^\rho \frac{k\hspace{-.45em}/\, + \Sigma (k^2)}{k^2 -\Sigma^2 (k^2)}
\Gamma^\mu_5 (k)
\frac{k\hspace{-.45em}/\, + \Sigma (k^2)}{k^2 -\Sigma^2 (k^2)} \gamma^\lambda 
,
\nonumber\\
\label{eq:axialsde}
\end{eqnarray}
where $ D_{\rho \lambda} (q) \equiv \frac{-1}{q^2} \left( g_{\rho \lambda} - \frac{q_\rho q_\lambda }{q^2} \right) $ is the gluon propagator in the Landau gauge (the color index is factorized), and $\Gamma_5^\mu$ is the dynamical axial charge in the zero limit of the momentum transfer.

In Eq. (\ref{eq:axialsde}), there are three relevant Lorentz structures: $\gamma^\mu \gamma_5$, $i\sigma^{\mu \nu} p_\nu \gamma_5$, and $p^\mu p\hspace{-.45em}/\, \gamma_5$.
The dynamical axial charge is thus written as 
\begin{eqnarray}
\Gamma^\mu_5 (p) 
&\equiv &
G_1 (p^2) \gamma^\mu \gamma_5
+G_2 (p^2) i\sigma^{\mu \nu} p_\nu \gamma_5
\nonumber\\
&&
+G_3 (p^2) p^\mu p\hspace{-.45em}/\, \gamma_5
\, .
\label{eq:axialstructure}
\end{eqnarray}
The SDE (\ref{eq:axialsde}) can thus be rewritten as a set of integral equations with the $G_1 (p^2)$, $G_2 (p^2)$, and $G_3 (p^2)$ functions.
The zero momentum point of the $G_1$ function indicates the ratio between the axial charges of the dressed and bare quarks (it will simply be called the ``quark axial charge'' from now on).

\begin{figure}[htb]
\includegraphics[width=8cm]{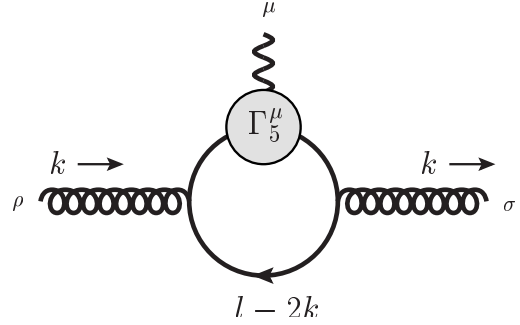}
\caption{\label{fig:axial_inner_loop}
Unquenching quark-loop diagram of the quark axial charge.
This diagram contributes to the chiral anomaly, so the loop momentum of the integral must be specified to be consistent with the vector Ward identity.
}
\end{figure}

\begin{figure}[htb]
\includegraphics[width=17cm,angle=-90]{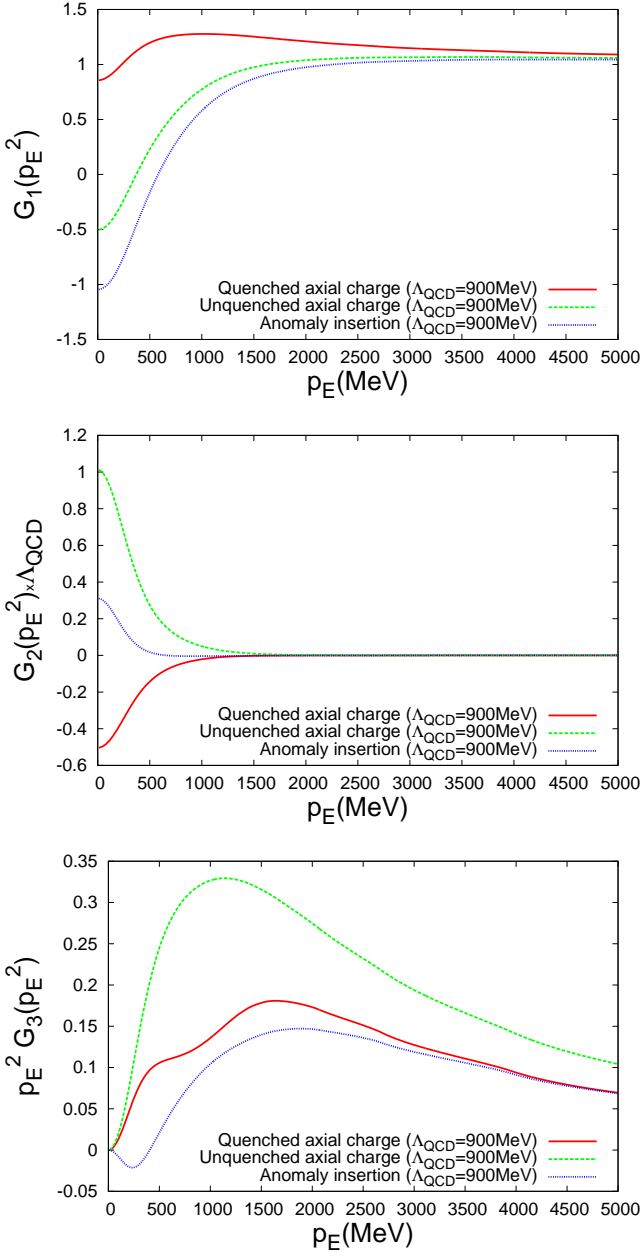}
\caption{\label{fig:axial_charge_123}
The $G_1$, $G_2$, and $G_3$ functions solved with the Schwinger-Dyson equation for the quark axial charge with the integral cutoff $\Lambda = 10$ GeV.
The $G_3$ function is resized with $p^2$.
The renormalization scale is taken as $\mu = 2$ GeV.
``Anomaly insertion'' corresponds to the solution of the quark axial SDE with the effect of the unquenching quark-loop replaced by the axial anomaly current.
}
\end{figure}

After some algebra, the quenched (isovector) axial SDE (\ref{eq:axialsde}), which does not include the two-loop level term (the last term of the right-hand side of Fig. \ref{fig:axial_SD_eq}) is given by the following set of integral equations:
\begin{widetext}
\begin{eqnarray}
G_1 (p_E^2) 
&=&
1
+ \frac{C_2 (N_c)}{3\pi^2} \int_0^\Lambda \hspace{-0.5em} k_E^3d k_E \, \int_0^\pi \hspace{-0.3em} \sin^2 \theta d\theta
 \frac{\alpha_s \left[ (p_E-k_E)^2 \right] Z^2 (k_E^2)}{\left[ k_E^2 + \Sigma^2 (k_E^2) \right]^2 p_E^2}
\nonumber\\
&& \hspace{6em} \times \Biggl\{ \ 
G_1 (k_E^2) \Biggl[ 
\frac{\Sigma^2 (k_E^2) -p_E^2}{2} \frac{(p_E^2 -k_E^2)^2}{(p_E-k_E)^4} 
+ \frac{2p_E^4 +2p_E^2k_E^2 +k_E^4 -\Sigma^2 (k_E^2) (4p_E^2 + k_E^2 ) }{ (p_E-k_E)^2} 
\nonumber\\
&& \hspace{24em}
-\frac{5p_E^2+4k_E^2 -\Sigma^2 (k_E^2) }{2} 
+(p_E-k_E)^2
\Biggr]
\nonumber\\
&& \hspace{7em} 
+G_2 (k_E^2) \Sigma (k_E^2) \Biggl[ 
-\frac{p_E^2 +k_E^2}{2} \frac{(p_E^2 -k_E^2)^2}{(p_E-k_E)^4} 
+ 2\frac{p_E^4 +3p_E^2 k_E^2 + k_E^4 }{ (p_E-k_E)^2} 
\nonumber\\
&& \hspace{24em}
-\frac{5}{2} (p_E^2 +k_E^2 )
+(p_E-k_E)^2
\Biggr]
\nonumber\\
&& \hspace{7em} 
+G_3 (k_E^2) \frac{k_E^2+ \Sigma^2 (k_E^2)}{2} \Biggl[ 
\frac{p_E^2-k_E^2}{2} \frac{(p_E^2 -k_E^2)^2}{(p_E-k_E)^4} 
+2 \frac{p_E^2 (k_E^2- p_E^2) }{ (p_E-k_E)^2} 
\nonumber\\
&& \hspace{24em}
+\frac{5p_E^2+3k_E^2}{2}
-(p_E-k_E)^2
 \Biggr]
\ \Biggr\} 
\ ,
\label{eq:sdeg1}
\\
G_2 (p_E^2)
&=&
\frac{C_2 (N_c)}{3\pi^2} \int_0^\Lambda \hspace{-0.5em} k_E^3d k_E \, \int_0^\pi \hspace{-0.3em} \sin^2 \theta d\theta
 \frac{\alpha_s \left[ (p_E-k_E)^2 \right] Z^2 (k_E^2)}{\left[ k_E^2 + \Sigma^2 (k_E^2) \right]^2 p_E^2}
\cdot \left[ 
2 \frac{(p_E^2 -k_E^2)^2}{(p_E-k_E)^4} - \frac{p_E^2+k_E^2 }{(p_E-k_E)^2} -1
\right]
\nonumber\\
&& \hspace{20em} \times \Biggl\{ \ 
G_1 (k_E^2) \Sigma (k_E^2) 
-G_2 (k_E^2) \frac{k_E^2 -\Sigma^2 (k_E^2)}{2}
\ \Biggr\} 
\ ,
\label{eq:sdeg2}
\end{eqnarray}
\begin{eqnarray}
G_3 (p_E^2) 
&=&
\frac{C_2 (N_c)}{3\pi^2} \int_0^\Lambda \hspace{-0.5em} k_E^3d k_E \, \int_0^\pi \hspace{-0.3em} \sin^2 \theta d\theta
 \frac{\alpha_s \left[ (p_E-k_E)^2 \right] Z^2 (k_E^2)}{\left[ k_E^2 + \Sigma^2 (k_E^2) \right]^2 p_E^4}
\nonumber\\
&& \hspace{6em} \times \Biggl\{ \ 
G_1 (k_E^2) \Biggl[ 
\left\{ p_E^2+2\Sigma^2 (k_E^2) \right\} \frac{(p_E^2 -k_E^2)^2}{(p_E-k_E)^4} 
+ 2 \frac{p_E^4 +p_E^2k_E^2 +2k_E^4 +\Sigma^2 (k_E^2) (p_E^2 -2 k_E^2 ) }{ (p_E-k_E)^2} 
\nonumber\\
&& \hspace{24em}
-7 p_E^2 -8 k_E^2 +2\Sigma^2 (k_E^2)
+4 (p_E-k_E)^2
\Biggr]
\nonumber\\
&& \hspace{7em} 
+G_2 (k_E^2) \Sigma (k_E^2) \Biggl[ 
( p_E^2 -2k_E^2) \frac{(p_E^2 -k_E^2)^2}{(p_E-k_E)^4} 
+ 2\frac{p_E^4 + 4 k_E^4 }{ (p_E-k_E)^2} 
-7p_E^2 -10k_E^2
+4(p_E-k_E)^2
\Biggr]
\nonumber\\
&& \hspace{7em} 
+G_3 (k_E^2) \frac{k_E^2+ \Sigma^2 (k_E^2)}{2} \Biggl[ 
-(p_E^2+2k_E^2) \frac{(p_E^2 -k_E^2)^2}{(p_E-k_E)^4} 
-2 \frac{p_E^2 (p_E^2 + 2k_E^2) }{ (p_E-k_E)^2} 
\nonumber\\
&& \hspace{26em}
+ 7p_E^2+6k_E^2
-4(p_E-k_E)^2
\Biggr]
\ \Biggr\} 
\ ,
\label{eq:sdeg3}
\end{eqnarray}
\end{widetext}
where $(p_E -k_E)^2 \equiv p_E^2+k_E^2 -2 p_E k_E \cos \theta $.
For the derivation of the above integral equations, see Appendix \ref{sec:axial_derivation_1-loop}.
The result of the SDE for the quenched (isovector) quark axial charge is plotted in Fig. \ref{fig:axial_charge_123}.

We now include the unquenching (isoscalar) quark-loop effect (see the last diagram of Fig. \ref{fig:axial_SD_eq}).
This 2-loop diagram is an isoscalar contribution, so it has no effect on the isovector axial charge.
It is important to note that the inner quark-loop is the triangle diagram contribution of the chiral (Adler-Bell-Jackiw) anomaly (see Fig. \ref{fig:axial_inner_loop}) \cite{abj}.
Due to the linear divergence of the triangle anomaly diagram, the loop integral depends on the choice of the shift of the loop momentum, and this momentum shift should be determined so as to fulfill the vector Ward identity \cite{cheng-li}.
In this work, we have chosen the shift of the momentum so that the vector Ward identity is realized in the infinite limit of the momentum cutoff $\Lambda \rightarrow \infty$ of the loop integral (see the momentum assignment of Fig. \ref{fig:axial_inner_loop}).
We also note that in our calculation, the effect of the strange quark-loop was neglected.

The unquenched (isoscalar) axial SDE is given by
\begin{widetext}
\begin{eqnarray}
G_1 (p_E^2) 
&=&
-\frac{C_2(N_c)}{3 \pi^2 p_E^2} \int_0^\Lambda \hspace{-0.5em} dk_E \frac{\alpha_s^2 (k_E^2)}{k_E} \int_0^\pi \hspace{-0.5em} \sin^2 \theta d \theta
\frac{Z[(p_E -k_E)^2]}{(p_E -k_E )^2 +\Sigma^2 [(p_E -k_E )^2]} f_5 \bigl[G_1,G_2,G_3 ; k_E^2 \bigr]
\nonumber\\
&& \hspace{4em}
\times \Biggl[ 
\frac{5}{2} p_E^4 -2 p_E^2 k_E^2 -\frac{1}{2} k_E^4 +(k_E^2 -2p_E^2)(p_E-k_E)^2 -\frac{1}{2} (p_E-k_E)^4
\Biggr]
+[\mbox{RHS of Eq. (\ref{eq:sdeg1})}]
, \ \ 
\label{eq:sdeg1_isoscalar}
\\
G_2 (p_E^2) 
&=&
-\frac{C_2(N_c)}{3 \pi^2 p_E^2} \int_0^\Lambda \hspace{-0.5em} dk_E \frac{\alpha_s^2 (k_E^2)}{k_E} \int_0^\pi \hspace{-0.5em} \sin^2 \theta d \theta
\frac{Z[(p_E -k_E)^2]}{(p_E -k_E )^2 +\Sigma^2 [(p_E -k_E )^2]} f_5 \bigl[G_1,G_2,G_3 ; k_E^2 \bigr]
\nonumber\\
&& \hspace{12em}
\times 3 \Sigma \bigl[ (p_E-k_E)^2 \bigr] \,
\Biggl[ p_E^2 + k_E^2 -(p_E -k_E)^2 \Biggr]
+[\mbox{RHS of Eq. (\ref{eq:sdeg2})}]
,
\label{eq:sdeg2_isoscalar}
\\
G_3 (p_E^2) 
&=&
-\frac{C_2(N_c)}{3 \pi^2 p_E^4} \int_0^\Lambda \hspace{-0.5em} dk_E \frac{\alpha_s^2 (k_E^2)}{k_E} \int_0^\pi \hspace{-0.5em} \sin^2 \theta d\theta
\frac{Z[(p_E -k_E)^2]}{(p_E -k_E )^2 +\Sigma^2 [(p_E -k_E )^2]} f_5 \bigl[G_1,G_2,G_3 ; k_E^2 \bigr]
\nonumber\\
&& \hspace{4em}
\times \Biggl[ 
p_E^4 + p_E^2 k_E^2 -2 k_E^4 +(p_E^2 +4k_E^2)(p_E-k_E)^2 -2 (p_E-k_E)^4
\Biggr]
+[\mbox{RHS of Eq. (\ref{eq:sdeg2})}]
,
\label{eq:sdeg3_isoscalar}
\end{eqnarray}
where the function $f_5 \bigl[G_1,G_2,G_3 ; k_E^2 \bigr]$ is defined as
\begin{eqnarray}
f_5 \bigl[G_1,G_2,G_3 ; k_E^2 \bigr]
&=&
\frac{2N_f}{\pi^2 k_E^2} \int_0^\Lambda \hspace{-0.5em} l_E^3 dl_E \int_0^\pi \hspace{-0.5em} \sin^2 \theta d\theta
\frac{Z^2 [(l_E -k_E)^2]\, Z[(l_E -2k_E)^2]}{\Bigl[ (l_E -2k_E )^2 +\Sigma^2 [(l_E -2k_E)^2] \Bigr] \Bigl[ (l_E -k_E)^2 +\Sigma^2 [(l_E -k_E)^2] \Bigr]^2 } 
\nonumber\\
&& 
\times \Biggl\{
\ G_1 [(l_E-k_E)^2] 
\Biggl[ 
\frac{1}{3} \Bigl[ -(l_E^2 -k_E^2)^2 +(5l_E^2 -7k_E^2 )(l_E-k_E)^2 -4(l_E -k_E)^4 \Bigr]
\nonumber\\
&& \hspace{9em}
+\Sigma^2 [(l_E-k_E)^2]
\Bigl[ 3k_E^2 -l_E^2 + (l_E-k_E)^2 \Bigr]
\nonumber\\
&& \hspace{9em}
+2\Sigma \bigl[ (l_E-2k_E)^2 \bigr] \Sigma [(l_E-k_E)^2] \Bigl[ l_E^2 - k_E^2 - (l_E-k_E)^2 \Bigr]
\Biggr]
\nonumber\\
&& \hspace{1.5em}
+G_2 [(l_E-k_E)^2]
\Biggl[ 
\frac{1}{3} \Sigma [(l_E-k_E)^2] 
\Bigl[
-(l_E^2 -k_E^2)^2 +8(l_E^2-2k_E^2)(l_E-k_E)^2-7(l_E-k_E)^4
\Bigr]
\nonumber\\
&& \hspace{9em}
+\Sigma \bigl[ (l_E-2k_E)^2 \bigr] 
\Bigl[ (l_E-k_E)^2 - \Sigma^2 [(l_E-k_E)^2] \Bigr]
\Bigl[ k_E^2 -l_E^2 + (l_E-k_E)^2 \Bigr]
\Biggr]
\nonumber\\
&& \hspace{1.5em}
+G_3 [(l_E-k_E)^2]
\frac{(l_E-k_E)^2 + \Sigma^2 [(l_E-k_E)^2]}{6} 
\nonumber\\
&& \hspace{10em} \times
\Bigl[
(l_E^2 -k_E^2)^2 -2(l_E^2 +k_E^2)(l_E-k_E)^2 +(l_E -k_E)^4
\Bigr]
\ \ \Biggr\}
. \ \ \ \ \ \ 
\label{eq:f_5euclid}
\end{eqnarray}
\end{widetext}
For the derivation of the above formula, see Appendix \ref{sec:axial_derivation_2-loop}.
Note that due to the omission of the strange quark-loop, $N_f =2$, but this number is only valid for the above equation.
The result of this unquenched (isoscalar) SDE of the quark axial charge is also plotted in Fig. \ref{fig:axial_charge_123}.

\subsection{\label{sec:axialanalysis}Quark axial charge: Analysis}

The quenched (isovector) quark axial charge SDE (\ref{eq:sdeg1}), (\ref{eq:sdeg2}), and (\ref{eq:sdeg3}) gives the following $G_1 (0)$
\begin{equation}
G_1 (0) = 0.86 \, ,
\label{eq:quenchedg_1}
\end{equation}
We see that $G_1$(0) is smaller than 1.
This fact shows that the quenched (isovector) axial charge of the dressed quark is suppressed compared with the bare quark contribution by the gluon dressing of the vertex.
We should note that an additional factor of renormalization is not needed for the quenched axial charge at the leading order of perturbation \cite{kodaira,QCDhigh}. 
By combining the above result with the isovector axial coupling predicted in the nonrelativistic constituent quark model without spin-dependent interactions ($g_A = \frac{5}{3}$), we obtain
\begin{equation}
g_A = 1.43 \, .
\label{eq:ourg_a}
\end{equation}
Here we have associated the dressed dynamical quark of the SD formalism with the massive constituent quark.
This manipulation has also been used for the estimation of the quark tensor charge \cite{tensorsde}.
Qualitatively, the result of Eq. (\ref{eq:ourg_a}) is in agreement with the experimental value (\ref{eq:isovectoraxialexp}).
However, we also observe some discrepancy between them. 
This shows that there are also other remaining effects besides the vertex gluon dressing which suppress the isovector axial charge of the nucleon.
One of the main candidates is the spin-dependent interactions of the quarks.

\begin{figure}[htb]
\includegraphics[width=6.1cm,angle=-90]{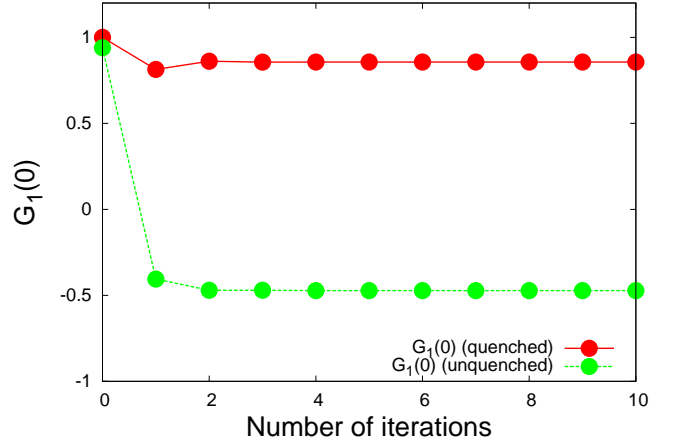}
\caption{\label{fig:axial_a1(0)}
The convergence of the renormalized $G_1$ function at the origin vs the number of iterations of the Schwinger-Dyson equation with the initial conditions $G_1(p_E^2) = 1$, $G_2(p_E^2) = 0$, and $G_3(p_E^2) = 0$.
The initial value $G_1(p_E^2) = 1$ is the bare quark axial charge.
}
\end{figure}

\begin{figure}[htb]
\includegraphics[width=8.6cm]{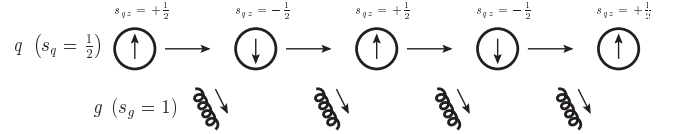}
\caption{\label{fig:quark_spin}
The schematic picture of the quark spin flip with the gluon emission.
The spins of the quark and the gluon are, respectively, $s_q=\frac{1}{2}$ and $s_g=1$
}
\end{figure}

The explanation of the suppression of the single quenched (isovector) quark axial charge is similar to the mechanism of the suppression of the single quark tensor charge \cite{tensorsde}.
We now look at the quark axial charge obtained after a few iterations.
The quark axial charge $G_1 (0)$ calculated after each iteration is shown in Fig. \ref{fig:axial_a1(0)}.
In our calculation of the SDE, we have taken as the initial condition $G_1(p^2) = 1$, $G_2(p^2) = 0$, and $G_3(p^2) = 0$, and iteratively substituted the left-hand sides of Eqs. (\ref{eq:sdeg1}), (\ref{eq:sdeg2}), and (\ref{eq:sdeg3}) into their right-hand sides.
As we have seen for the quark scalar density, this procedure can be seen as a sort of perturbative truncation.
From Fig. \ref{fig:axial_a1(0)}, we can see that the quenched (isovector) axial charge converges by oscillating.
This means that the gluon-dressed axial vertex is decomposed into terms which alternate in sign in the perturbative expansion.
The calculation of the quark tensor charge exhibits a similar behavior \cite{tensorsde}.
This result shows that the reversal of the quark spin is preferred in gluon emission/absorption.
This is also consistent with the angular momentum conservation since the gluons have spin 1.
The above description is illustrated schematically in Fig. \ref{fig:quark_spin}.
Note that the helicity of the quark is not changed in the gluon emission and absorption.
As the external field can only probe the axial charge (spin) of the quark, the superposition of the contribution of each order is always smaller than the bare one.
Although the tensor and axial charges both give the quark spin in the nonrelativistic limit, their dynamical suppression factors are quite different [the quark tensor charge is suppressed by a factor of 0.6 (at the renormalization scale $\mu = 2 $ GeV)] \cite{tensorsde}.
Let us remember that the tensor charge is a chiral odd quantity, whereas the axial charge is chiral-even, so that their difference signals how relativistic the quark is \cite{tensornonrela}.
As the dressed quarks are not fully nonrelativistic (this fact can also be seen in the difference between the quark scalar and vector charges), this difference is quite natural.

We now discuss the unquenched (isoscalar) quark axial charge SDE (\ref{eq:sdeg1_isoscalar}), (\ref{eq:sdeg2_isoscalar}), and (\ref{eq:sdeg3_isoscalar}).
For the unquenched singlet axial charge, we need to pay attention to the renormalization.
The unquenching diagram we are treating involves the Adler-Bell-Jackiw triangle anomaly graph \cite{abj} (see Fig. \ref{fig:axial_inner_loop}), and this contribution needs to be renormalized \cite{kodaira,QCDhigh}.
At the scale $\mu$, the multiplicative renormalization of the singlet quark axial charge is given as \cite{QCDhigh}
\begin{equation}
G_1 (0)_\mu 
=
G_1 (0)_\Lambda
\times e^{ \frac{C_2 (N_c)}{4\pi \beta_0} (11 N_c - 8 N_f) [ \alpha_s (\Lambda^2 ) - \alpha_s (\mu^2 ) ] }
\, .
\end{equation}
where $G_1 (0)_\Lambda$ denotes the bare singlet quark axial charge.
The coefficient of the exponent is $ \frac{C_2 (N_c)}{4\pi \beta_0} (11 N_c - 8 N_f) = \frac{1}{3\pi}$ for $N_c = N_f =3$, and $ \frac{C_2 (N_c)}{4\pi \beta_0} (11 N_c - 8 N_f) = \frac{17}{29\pi}$ for $N_c =3, N_f =2$.
The multiplicative renormalization $e^{ \frac{C_2 (N_c)}{4\pi \beta_0} (11 N_c - 8 N_f) [ \alpha_s (\Lambda^2 ) - \alpha_s (\mu^2 ) ] } $ is close to 1, so the effect of the renormalization is not important.

The result of the singlet unquenched quark axial charge renormalized at $\mu = 2$ GeV gives
\begin{equation}
G_1 (0)_{\mu = 2 \, {\rm GeV}} = -0.47 \, .
\label{eq:unquenchedg_1}
\end{equation}
We see that $G_1$(0) is much smaller than the quenched case (\ref{eq:quenchedg_1}), and it is well below zero [see also Fig. \ref{fig:axial_charge_123}].
This result suggests that the axial anomaly has a significant effect on the suppression of the single isoscalar quark axial charge.
As we have remarked in the previous section, the inner quark-loop of the unquenching effect in the quark axial charge SDE (Fig. \ref{fig:axial_SD_eq}) is due to the axial anomaly.
We should note that, in our calculation, the integral of the inner loop was also cut off at $\Lambda = 10$ GeV.
To reproduce the axial anomaly, we must integrate the inner loop integral function $f_5$ [see Eq. (\ref{eq:f_5euclid})] with the cutoff $\Lambda \rightarrow \infty$ with $G_1 [(l_E-k_E)^2] =1$, $G_2 [(l_E-k_E)^2] = G_3 [(l_E-k_E)^2] =0$.
We therefore obtain
\begin{equation}
f_5 [1,0,0,0;k_E^2] =-\frac{N_f}{\pi} 
\, ,
\end{equation}
which is the exact axial anomaly contribution.
We show also in Fig. \ref{fig:axial_charge_123} the result obtained after the insertion of the bare axial anomaly instead of the unquenching quark-loop.
It can be seen that the bare anomaly effect is larger than the quark-loop contribution of our calculation.
This difference should be due to the low cutoff $\Lambda =10$ GeV we have used in the inner loop integration.
The unquenching quark-loop effect should approach the bare anomaly when the cutoff is enlarged, but this requires a large computational effort.

We should also note that the transfer of the quark spin to the orbital angular momentum may be an additional source of the suppression of the quark axial charge.
In our framework, it is not possible to distinguish the effect of the anomaly from that of the orbital angular momentum.
The formulation of the angular momenta of quarks and gluons has recently seen much development \cite{angularmomentum}.
To study this effect, we must evaluate the insertion of the orbital angular momentum operator into the unquenching quark-loop diagram in the SD formalism, but this work is beyond the scope of this paper.

In our formalism, we have obtained a large suppression of the quark axial charge due to the unquenching quark-loop diagram.
We should however note that this result was obtained in a framework that does not consider quark confinement.
As we have seen for the quark scalar density, the unquenching quark-loop effect may be significantly suppressed by the reduction of the configurations of the path of the quarks by the confining potential.
A quantitative evaluation of the quark axial charge needs a careful treatment of the IR region.
In evaluating the nucleon axial charge, we can also expect a sizable contribution from the exchanged current due to the quark-loops (see Fig. \ref{fig:exchangecurrent}), which is not necessarily of the same sign as the quark-loop contributing to the axial charge of the single quark.
Phenomenologically, it can be estimated that the effect of the singlet axial anomaly on the proton spin is not very large \cite{cheng-li_prl}.
It will thus be important to compare the effect of the axial anomaly from the single quark with that of the many-body interactions to determine the source of the proton spin crisis.
The development of the Nambu-Goldstone mode as a higher order unquenching effect cannot be neglected either \cite{unquenching}.

We also add some comments on the dependence of the quark axial charge on the scale parameter $\Lambda_{\rm QCD}$.
We show the coefficient $G_1 (0)$ for several values of $\Lambda_{\rm QCD}$ in Table \ref{table:axiallambdaqcddependence}.
For the quenched axial charge, the dependence is small. 
This stable behavior is similar to that of the tensor charge \cite{tensorsde}. 
\begin{table}
\caption{The quark axial charge obtained with several values of $\Lambda_{\rm QCD}$.
The renormalization scale was fixed to $\mu = 2$ GeV.
}
\begin{ruledtabular}
\begin{tabular}{ccccc}
$\Lambda_{\rm QCD}$ & 200 MeV & 500 MeV & 900 MeV & 1.3 GeV \\
\hline
quenched & 0.863 & 0.858 & 0.857 & 0.856 \\
unquenched & -0.626 & -0.568 & -0.473 & -0.366 \\
bare anomaly &-1.036 &-1.023 &-0.982 & -0.906\\
\end{tabular}
\end{ruledtabular}
\label{table:axiallambdaqcddependence}
\end{table}
For the unquenched case, however, we can see some dependence on the QCD scale parameter.
There the deviation of the quark axial charge from 1 becomes smaller for large values of $\Lambda_{\rm QCD}$.
This can be explained by the fact that the quark-loop effect becomes larger when the quark has a smaller dynamical mass.
The unquenched effect due to the bare anomaly also becomes smaller for large values of $\Lambda_{\rm QCD}$, but the variation is not as significant as the quark-loop unquenching contribution.

\begin{figure}[htb]
\includegraphics[width=6.1cm,angle=-90]{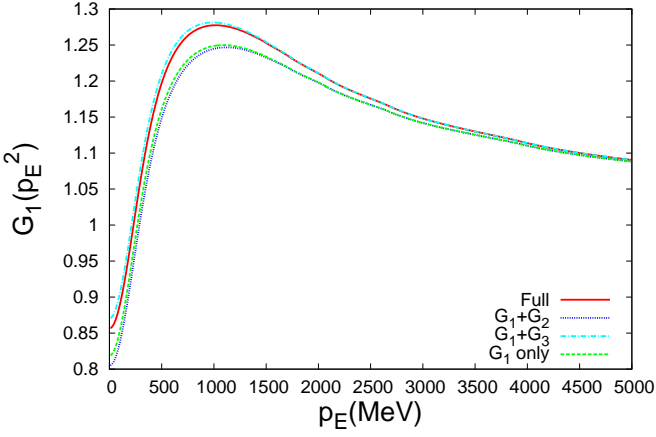}
\caption{\label{fig:axial_a1_compare}
The $G_1$ function obtained by solving the Schwinger-Dyson equation with the $G_2$ and $G_3$ functions set to zero.
The $G_1$ function solved with the full contribution ($G_1$, $G_2$, and $G_3$) is also shown for comparison.
}
\end{figure}

In the case of the quenched (isovector) quark axial charge, it is also possible to approximately reduce the SDE.
Again, we perform a fictitious manipulation by setting $G_2 (p^2) =0$ and/or $G_3 (p^2)=0$ when solving the SDE (\ref{eq:sdeg1}), (\ref{eq:sdeg2}), and (\ref{eq:sdeg3}).
The result is plotted in Fig. \ref{fig:axial_a1_compare}.
We see that the solutions of the SDE with and without the contribution from the $G_2$ and $G_3$ functions are close and the qualitative features are very similar.
This shows that the leading contribution to the SDE of the quenched quark axial charge is given by the $G_1$ function, and that the omission of $G_2$ and $G_3$ is a relatively good approximation.

It should be noted that this approximative reduction does not work for the unquenched (isoscalar) quark axial charge SDE.
It can be inferred that the momentum dependence of the dynamical axial charge ($G_2$ and $G_3$) plays an important role in the effective vertex of closed quark-loops.

\section{\label{sec:pseudoscalar}Quark pseudoscalar density}

\begin{figure}[htb]
\begin{center}
\includegraphics[width=8cm]{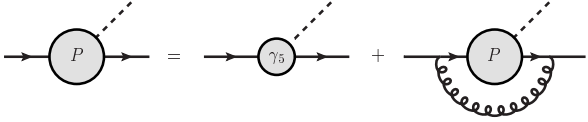}
\caption{\label{fig:pseudoscalar_SD_eq}
Diagrammatic picture of the Schwinger-Dyson equation for the quark pseudoscalar density.
}
\end{center}
\end{figure}

The Schwinger-Dyson equation for the quark pseudoscalar density is given by
\begin{eqnarray}
P (p^2)\gamma_5
&=&
\gamma_5
\nonumber\\
&&
+
i C_2 (N_c) \int \frac{d^4k}{4\pi^3}
\alpha_s \left[(p-k)^2 \right] Z^2 (k^2)
\nonumber\\
&& \times
\gamma^\rho \frac{k\hspace{-.45em}/\, + \Sigma (k^2)}{k^2 -\Sigma^2 (k^2)}
P (k^2) \gamma_5
\frac{k\hspace{-.45em}/\, + \Sigma (k^2)}{k^2 -\Sigma^2 (k^2)} \gamma^\lambda 
\nonumber\\
&&\times D_{\rho \lambda} (p-k)
,
\label{eq:pseudoscalarsde}
\end{eqnarray}
where $P$ is the dynamical pseudoscalar density, $Z$ is the quark wave function renormalization, $\Sigma$ is the quark self-energy, and $\alpha_s [(p-k)^2]$ is the RG-improved strong coupling.
The Schwinger-Dyson equation for the pseudoscalar density is depicted diagrammatically in Fig. \ref{fig:pseudoscalar_SD_eq}.

\begin{figure}[htb]
\includegraphics[width=5cm]{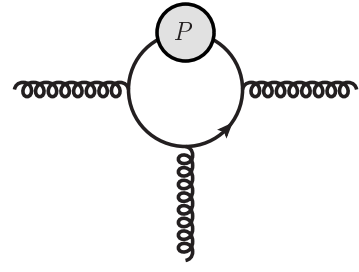}
\caption{\label{fig:3-gluon}
quark-loop diagram coupled to three gluons.
The grey blob represents the pseudoscalar operator insertion.
This contribution is not vanishing.
}
\end{figure}

After Wick rotation, Eq. (\ref{eq:pseudoscalarsde}) can be rewritten as
\begin{eqnarray}
P (p_E^2)
&=&
1+
\frac{3 C_2 (N_c)}{\pi^2} 
\int_0^\infty \hspace{-0.5em} k_E^3 dk_E \int_0^\pi \hspace{-0.3em} \sin^2 \theta d\theta 
\nonumber\\
&&\times
\frac{\alpha_s ( p_E^2+k_E^2 -2p_E k_E \cos \theta ) \, Z^2 (k_E) P (k_E) }{( p_E^2+k_E^2 -2p_E k_E \cos \theta ) \left[ k_E^2 +\Sigma^2 (k_E^2) \right]}
 .
\nonumber\\
\label{eq:pseudoscalarsde}
\end{eqnarray}
It should be noted that the pseudoscalar density receives no corrections from the unquenching quark-loop with two gluons (see the last term of the SDE of Figs. \ref{fig:scalar_SD_eq} and \ref{fig:axial_SD_eq}) even for the isoscalar contribution.
This is because the pseudoscalar insertion in the inner quark-loop does not have sufficient gamma matrices to obtain a nonzero Dirac trace.
This does not mean that the quark pseudoscalar density has no unquenching quark-loop effect, since the quark-loop with a pseudoscalar insertion and three gluons (see Fig. \ref{fig:3-gluon}) is nonzero.
This can be shown with Furry's theorem \cite{furry}.

\begin{figure}[htb]
\includegraphics[width=6.1cm,angle=-90]{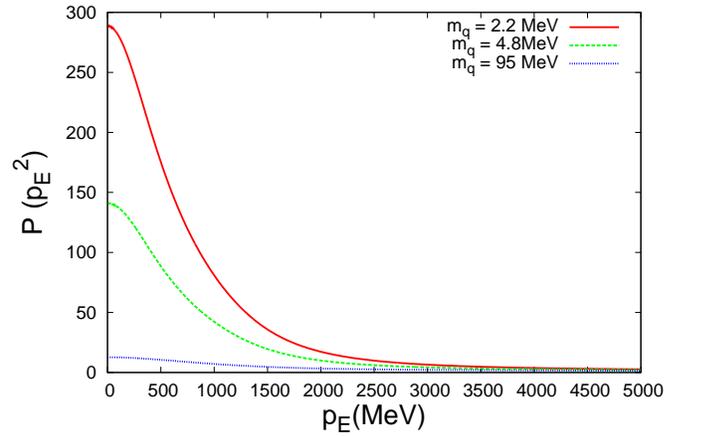}
\caption{\label{fig:pseudoscalar_charge}
The $P$ function (not renormalized) obtained by solving the Schwinger-Dyson equation with different current quark masses.
}
\end{figure}

\begin{figure}[htb]
\includegraphics[width=6.1cm,angle=-90]{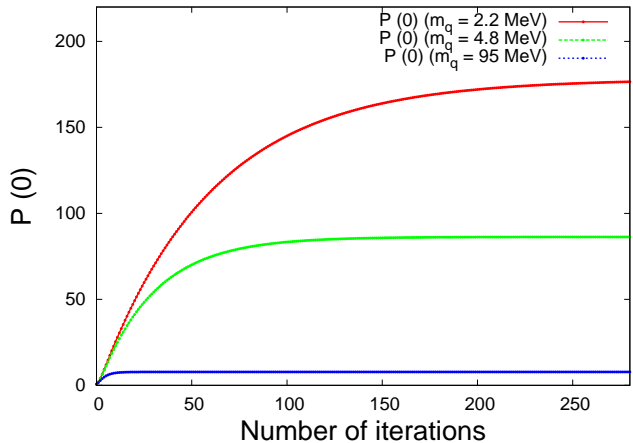}
\caption{\label{fig:pseudoscalar_p_0}
The convergence of the renormalized $P$ function at $p_E^2 =0$ vs the number of iterations of the Schwinger-Dyson equation with the initial conditions $P(p_E^2) = 1$ for different current quark masses.
The initial value $P(p_E^2) = 1$ is the bare quark pseudoscalar density.
}
\end{figure}

We have solved the quark pseudoscalar density SDE with several quark masses, $m_q = 2.2$ MeV, 4.8 MeV, and 95 MeV, corresponding to the mass of the up, down, and strange quarks, respectively, at the renormalization point $\mu=2$ GeV \cite{pdg}.
The result of the calculation is plotted in Fig. \ref{fig:pseudoscalar_charge}.
The renormalization of the pseudoscalar density works similarly as for the scalar density (\ref{eq:scalar_density_renormalization}), 
\begin{equation}
P (0)_\mu 
=
\left( \frac{\alpha_s (\Lambda^2 )}{\alpha_s (\mu^2 ) } \right)^{-\frac{3C_2 (N_c)}{\beta_0}}
P (0)_\Lambda
\, ,
\label{eq:pseudoscalar_density_renormalization}
\end{equation}
where $P (0)_\Lambda$ is the quark pseudoscalar density obtained after solving the SDE (\ref{eq:pseudoscalarsde}) with the integral cutoff $\Lambda$.
We therefore obtain
\begin{eqnarray}
P(0)_{\mu =2\, {\rm GeV}}  
&=& 
177
\ \ \ (m_q = 2.2 \, {\rm MeV})
,
\nonumber\\
P(0)_{\mu =2\, {\rm GeV}}  
&=& 
86.3
\ \ \ (m_q = 4.8 \, {\rm MeV})
,
\nonumber\\
P(0)_{\mu =2\, {\rm GeV}}  
&=& 
7.76
\ \ \ (m_q = 95 \, {\rm MeV})
.
\end{eqnarray}
We see that the quark pseudoscalar density becomes larger for lighter quarks.
For the chiral limit, the quark pseudoscalar SDE (\ref{eq:pseudoscalarsde}) does not converge.
The convergence of the quark pseudoscalar density SDE is shown in Fig. \ref{fig:pseudoscalar_p_0}.
We see that the convergence is rather slow.

\begin{figure}[htb]
\includegraphics[width=8cm]{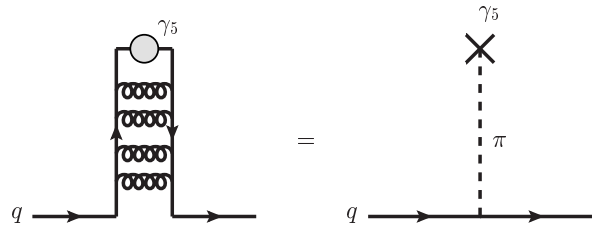}
\caption{\label{fig:pionpole}
The pion-pole contribution is generated by the ladder diagram.
}
\end{figure}

Let us now try to explain the large value of the pseudoscalar density of the light quarks.
Phenomenologically, the pseudoscalar content of the nucleon $\langle N | \bar q i \gamma_5 q | N \rangle$ is known to be large, due to the pion-pole contribution (see Fig. \ref{fig:pionpole}) \cite{cheng-li_prl,edm2,herczeg2,isovectorpseudoscalardensity}.
This can be estimated phenomenologically as
\begin{eqnarray}
\langle N | \bar q i \gamma_5 q | N \rangle
&\propto &
g_{\pi NN} \frac{1}{m_\pi^2} \langle 0 | \bar q i \gamma_5 q | \pi \rangle
\nonumber\\
&\sim &
g_{\pi NN} \frac{1}{f_\pi m_\pi^2} \langle 0 | \bar q q | 0 \rangle
\nonumber\\
&\sim &
O (100)
\, ,
\label{eq:pseudoscalardensitypheno}
\end{eqnarray}
where we have used $g_{\pi NN} \approx 13$, the PCAC reduction, and the Gell-Mann-Oakes-Renner relation.
This large value can make the observable effects of the quark pseudoscalar density important, although $\langle N | \bar q i \gamma_5 q | N \rangle$ is suppressed nonrelativistically \cite{herczeg2,isovectorpseudoscalardensity}.
The expression of Eq. (\ref{eq:pseudoscalardensitypheno}) is divergent in the chiral limit $m_q \rightarrow 0$, and this fact explains the large value of the matrix element $\langle N | \bar q i \gamma_5 q | N \rangle$ for light quarks.

This pion-pole effect must be relevant in our formulation of the single quark pseudoscalar density, since the interacting quark-antiquark pair as shown in Fig. \ref{fig:pionpole} generates a massless Nambu-Goldstone mode in the SD formalism through the ladder approximation \cite{pi-k,kaoki}.
We should note that this massless mode appears in any choice of the phenomenological interquark potential since we respect the chiral symmetry of the lagrangian and this symmetry is broken spontaneously.
This is one of the advantages of the SD formalism. 
In some sense, we may say that the quark pseudoscalar density is an observable sensitive to the light quark mass.

\section{\label{sec:summary}Summary}

In this paper, we have studied the quark scalar density, the quark axial charge, and the quark pseudoscalar density in the SD formalism of Landau gauge QCD.

For the quark scalar density, the result of our calculation shows an enhancement from the bare one in the quenched study.
The physical meaning of the quark scalar density is the sum of the probability of finding a quark in the whole space-time of the intermediate state.
This has been confirmed by comparing the quark scalar density calculated with and without the quark confinement effect, using the phenomenological Richardson Ansatz and the Higashijima-Miransky approximation.
We have also given the analytic formula for the leading unquenching effect on the quark scalar density, but the SDE does not converge with our setup.
This failure may be explained by the fact that the configuration of the quark-loop is too large, as we have ignored the effect of quark confinement.

For the quark axial charge, we have shown that it is suppressed by the gluon emission/absorption in the quenched case, for a reason similar to that found in the study of the quark tensor charge \cite{tensorsde}.
The quark axial charge receives a larger suppression by including the unquenching quark-loop effect.
This is due to the Adler-Bell Jackiw triangle anomaly, but there may also be an additional source of suppression due to the transfer of the quark spin to the orbital angular momentum.
This suppression may be determined quantitatively in a future work by analyzing the insertion of the quark orbital angular momentum operator into the unquenching quark-loop diagram in the SD formalism.
This unquenching effect may also be largely overestimated due to the large uncertainty in treating the IR region.
The axial anomaly also contributes to the many-body effect via the exchange interaction.
To study the problem of the proton spin quantitatively, we must therefore evaluate the many-body effect together with the discussion of this paper.

Finally, we have derived the analytic formula and calculated the quark pseudoscalar density in the SD formalism.
For the pseudoscalar charge there are no unquenching quark-loop diagrams with two-gluon exchange, but the quark-loop is not forbidden beyond the three-gluon exchange contribution.
As a result, we have obtained a large pseudoscalar density for light quarks, and this result is consistent with previous phenomenological analyses.
The divergence of the quark pseudoscalar SDE in the chiral limit is a rather natural result, since the dominant contribution to the quark pseudoscalar density is given by the pion-pole which diverges with massless quarks.
We conclude that the quark pseudoscalar density is an observable sensitive to the light quark mass.
The nonrelativistic suppression of the quark pseudoscalar density in the nucleon is compensated by its large value, and this may help the search for new particles beyond the standard model.

In this paper, we have also predicted that the hadronic molecules should have a larger quark scalar density than the single multi-quark baryons.
The quark scalar density is thus an observable sensitive to the compositeness of the hadrons, and this provides another qualitative way to approach the structure of hadrons.

We must however note that we have only discussed the single quark contribution to the nucleon charges.
The remaining effects to the nucleon charges should be investigated from the viewpoint of the many-body physics of partons.
This study will be the subject of our next work.
Here we briefly give the prospect for the improvement.
The first task is to improve the SD formalism by further considering the gluon sector and quark-gluon vertex SDE.
To obtain a more quantitative result, the inclusion of the effects beyond the rainbow-ladder approximation and a more sophisticated unquenching calculation will also be needed.
The ideal way to discuss the quark charges of the hadrons in the SD formalism is to formulate and calculate the relativistic Faddeev equation for the multi-quark states \cite{faddeev}.

\begin{acknowledgments}

NY thanks Y. Hatta, T. Hyodo, and H. Iida for useful discussions and comments.
This work is in part supported by the Grant for Scientific Research [Priority Areas ``New Hadrons'' (E01:21105006), (C) No.23540306] from the Ministry of Education, Culture, Science and Technology (MEXT) of Japan.

\end{acknowledgments}

\appendix

\onecolumngrid

\section{\label{sec:scalar_derivation}Detailed calculation of the Schwinger-Dyson equation for the quark scalar density}

\subsection{\label{sec:scalar_derivation_1-loop} Quark scalar density SDE: vertex dressing (quenched SDE)}

The Schwinger-Dyson equation for the quenched quark scalar density [Eq. (\ref{eq:scalarsde})] is rewritten as
\begin{eqnarray}
S_1 (p^2) 
+S_2 (p^2) p\hspace{-.45em}/\, 
&=&
1
+
i C_2 (N_c) \int \frac{d^4k}{4\pi^3} 
\cdot \frac{\alpha_s [(p-k)^2] Z^2 (k^2)}{\left[ k^2 -\Sigma^2 (k^2) \right]^2} 
\cdot \frac{-1}{(p-k)^2} \left[ g^{\rho \lambda} - \frac{(p-k)^\rho (p-k)^\lambda}{(p-k)^2} \right]
\nonumber\\
&& \hspace{8em} \times
\gamma^\rho 
\left[ k\hspace{-.45em}/\, + \Sigma (k^2)\right]
\left[ 
S_1 (k^2) 
+S_2 (k^2) k\hspace{-.45em}/\, 
\right]
\left[ k\hspace{-.45em}/\, + \Sigma (k^2)\right]
\gamma^\lambda 
.
\label{eq:scalarsde2}
\end{eqnarray}

The Lorentz and Dirac structures of the term with $S_1 (k^2)$ of Eq. (\ref{eq:scalarsde2}) can be transformed as
\begin{eqnarray}
\left[ g^{\rho \lambda} - \frac{(p-k)^\rho (p-k)^\lambda}{(p-k)^2} \right]
\gamma_\rho
\left[ k\hspace{-.45em}/\, + \Sigma \right]
\left[ k\hspace{-.45em}/\, + \Sigma \right] 
\gamma_\lambda
&=&
3\left[ k^2 + \Sigma^2  \right] 
+2\Sigma (p\hspace{-.45em}/\, -2k\hspace{-.45em}/\, )
-2\Sigma \frac{ p^2-k^2 }{(p-k)^2} (p\hspace{-.45em}/\, - k\hspace{-.4em}/\, )
\, .
\label{eq:s1lorentz}
\end{eqnarray}
For simplicity, we have omitted the argument of the self-energy $\Sigma$.
Similarly, the Lorentz and Dirac structures of the term with $S_2 (k^2)$ can be obtained as
\begin{eqnarray}
\left[ g^{\rho \lambda} - \frac{(p-k)^\rho (p-k)^\lambda}{(p-k)^2} \right]
\gamma_\rho
\left[ k\hspace{-.45em}/\, + \Sigma \right]
k\hspace{-.45em}/\,
\left[ k\hspace{-.45em}/\, + \Sigma \right] 
\gamma_\lambda
&=&
6 \Sigma k^2 
+( k^2 + \Sigma^2 ) (p\hspace{-.45em}/\, -2k\hspace{-.45em}/\, )
-( k^2 + \Sigma^2 ) \frac{ p^2-k^2 }{(p-k)^2} (p\hspace{-.45em}/\, - k\hspace{-.45em}/\, )
. \ \ \ \ 
\label{eq:s2lorentz}
\end{eqnarray}

By substituting Eqs. (\ref{eq:s1lorentz}) and (\ref{eq:s2lorentz}) into Eq. (\ref{eq:scalarsde2}), we can further transform the integral equation as
\begin{eqnarray}
S_1 (p^2) 
+S_2 (p^2) p\hspace{-.45em}/\, 
&=&
1
-i \int \frac{d^4k}{4\pi^3}
\cdot \frac{\alpha_s [(p-k)^2]Z^2 (k^2) }{\left[ k^2 -\Sigma^2 (k^2) \right]^2} 
\cdot \frac{C_2 (N_c)}{(p-k)^2} \cdot S_1(k^2)
\nonumber\\
&& \hspace{5em} \times 
\Biggl\{
3 \left[ k^2 + \Sigma^2 (k^2) \right]
- \Sigma (k^2) 
\left[ 
\frac{(p^2 -k^2)^2}{(p-k)^2}
+ p^2 +k^2
-2 (p-k)^2
\right]
\frac{p\hspace{-.45em}/ }{p^2}
\ \Biggr\}
\nonumber\\
&& \hspace{1em}
-i \int \frac{d^4k}{4\pi^3}
\cdot \frac{\alpha_s [(p-k)^2]Z^2 (k^2) }{\left[ k^2 -\Sigma^2 (k^2) \right]^2} 
\cdot \frac{C_2 (N_c)}{(p-k)^2} \cdot S_2(k^2)
\nonumber\\
&& \hspace{5em} \times 
\Biggl\{
6 k^2 \Sigma (k^2)
-\frac{1}{2} \left[ k^2 + \Sigma^2 (k^2) \right]
\cdot 
\left[
\frac{(p^2- k^2)^2}{(p-k)^2}
+p^2 + k^2
-2(p-k)^2
\right]
\frac{p\hspace{-.45em}/ }{p^2}
\, \Biggr\}
\, ,
\label{eq:scalarsde3}
\end{eqnarray}
which gives Eqs. (\ref{eq:sdes1}) and (\ref{eq:sdes2}).
Here we have used the formulae of the loop integral developed by Passarino and Veltman \cite{passarino-veltman} to reduce into a Lorentz scalar loop integral.
The rank-1 ($k^\mu$) integral can be reduced as
\begin{equation}
\int d^4 k \, F_1 (p,k) k^\mu
=
T_1 (p^2) p^\mu
\, ,
\end{equation}
where
\begin{equation}
T_1 (p^2) 
=
\int d^4 k \, F_1 (p, k ) \left[ \frac{p^2+k^2- (p-k)^2}{2p^2} \right]
=
\frac{1}{2}
\int d^4 k \, F_1 (p, k ) \left[ 1 +\frac{k^2}{p^2} - \frac{(p-k)^2}{p^2} \right]
\, .
\label{eq:t1}
\end{equation}
The rank-2 ($k^\mu k^\nu$) integral can be reduced as
\begin{equation}
\int d^4 k \, F_2 (p,k) k^\mu k^\nu
=
T_{00} (p^2) g^{\mu \nu}
+T_{11} (p^2) p^\mu p^\nu
\, ,
\end{equation}
where
\begin{eqnarray}
T_{00} (p^2) 
&=&
\frac{1}{3}
\int d^4 k \, F_2 (p,k) \, \left[ k^2 - \frac{1}{4} \frac{\left[ k^2 +p^2 -(p-k)^2 \right]^2}{p^2} \right]
\nonumber\\
&=&
\frac{1}{3}
\int d^4 k \, F_2 (p,k) \, \left[ \frac{1}{2} k^2 - \frac{1}{4} p^2 -\frac{1}{4} \frac{k^4}{p^2} +\frac{1}{2} \frac{(p^2+k^2)(p-k)^2}{p^2} -\frac{1}{4} \frac{(p-k)^4}{p^2} \right]
\, ,
\label{eq:t00}
\\
T_{11} (p^2) 
&=&
\frac{1}{3p^2}
\int d^4 k \, F_2 (p,k) \, \left[ \frac{\left[ k^2 +p^2 -(p-k)^2 \right]^2}{p^2} -k^2 \right]
\nonumber\\
&=&
\frac{1}{3}
\int d^4 k \, F_2 (p,k) \, \left[ 1+ \frac{k^2}{p^2} + \frac{k^4}{p^4} -2 \frac{(k^2+p^2)(p-k)^2}{p^4} +\frac{(p-k)^4}{p^4} \right]
\, .
\label{eq:t11}
\end{eqnarray}

\subsection{\label{sec:scalar_derivation_2-loop} Quark scalar density SDE: quark-loop contribution (unquenched isoscalar SDE)}

\begin{figure}[htb]
\includegraphics[width=8cm]{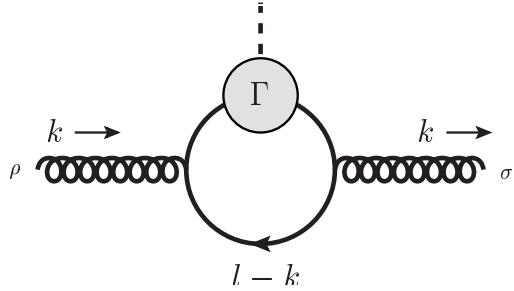}
\caption{\label{fig:scalar_inner_loop}
Inner quark-loop of the quark scalar charge diagram.
}
\end{figure}

To obtain the analytical expression of the quark-loop diagram contribution of the scalar SDE (\ref{eq:scalarsde}) (the last term of the right-hand side of the SDE  of Fig. \ref{fig:scalar_SD_eq}), we first calculate the inner quark-loop integral (see Fig. \ref{fig:scalar_inner_loop}).

The amplitude of the inner quark-loop is given as
\begin{equation}
i{\cal M}
=
-iN_f \frac{\alpha_s (k^2)}{4\pi^3} 
\epsilon^*_\sigma (k) \epsilon_\rho ( k)
\int d^4 l
\frac{{\rm Tr} \left[ \left\{ l\hspace{-.45em}/\, -k\hspace{-.45em}/\, +\Sigma [(l-k)^2] \right\} \gamma^\sigma \left\{ l\hspace{-.45em}/\, +\Sigma (l^2) \right\} \Gamma (l) \left\{ l\hspace{-.45em}/\, +\Sigma (l^2) \right\} \gamma^\rho \right] }{ \left\{ (l-k)^2 -\Sigma^2 [(l-k)^2] \right\} \left\{ l^2 - \Sigma^2 (l^2) \right\}^2 }
Z [(l-k)^2] Z^2 (l^2)
\delta_{ab}
\, ,
\label{eq:scalar_inner_loop}
\end{equation}
where $\Gamma (l) \equiv S_1 (l^2) +S_2 (l^2) l\hspace{-.45em}/\, $.
The color trace has already been reduced (${\rm tr}[t_a t_b ] = \frac{1}{2} \delta_{ab}$).
It is important to note that we have included in the above amplitude the effect of the quark propagating in the opposite direction.
The change of the direction of the quark propagation corresponds to the propagation of the antiquark.
It thus gives exactly the same contribution as for the quark-loop for the scalar density.
This can be easily shown by changing the integral variable $l\rightarrow -l$ and taking the transpose of the Dirac trace, where we use the charge conjugation property $C\gamma_\mu^T C^{-1} = -\gamma_\mu$.

Let us first consider the $S_1 $ contribution.
The trace is calculated as
\begin{eqnarray}
&&
{\rm Tr} \left[ \left\{ l\hspace{-.45em}/\, -k\hspace{-.45em}/\, +\Sigma [(l-k)^2] \right\} \gamma^\sigma \left\{ l\hspace{-.45em}/\, +\Sigma (l^2) \right\} \left\{ l\hspace{-.45em}/\, +\Sigma (l^2) \right\} \gamma^\rho \right] 
\nonumber\\
&=&
4\left[ l^2 +\Sigma^2 (l^2) \right] \Sigma [(l-k)^2] g^{\rho \sigma}
+8 \Sigma (l^2) \left[ 2l^\rho l^\sigma -l^2 g^{\rho \sigma} -l^\rho k^\sigma -k^\rho l^\sigma +(l\cdot k) g^{\rho \sigma} \right]
\, .
\end{eqnarray}
 
The $S_2$ contribution is calculated similarly as
\begin{eqnarray}
&&
{\rm Tr} \left[ \left\{ l\hspace{-.45em}/\, -k\hspace{-.45em}/\, +\Sigma [(l-k)^2] \right\} \gamma^\sigma \left\{ l\hspace{-.45em}/\, +\Sigma (l^2) \right\} l\hspace{-.45em}/\, \left\{ l\hspace{-.45em}/\, +\Sigma (l^2) \right\} \gamma^\rho \right] 
\nonumber\\
&=&
4\left[ l^2+ \Sigma^2 (l^2) \right] \left[ 2l^\rho l^\sigma -l^2 g^{\rho \sigma} -l^\rho k^\sigma -k^\rho l^\sigma +(l\cdot k) g^{\rho \sigma} \right]
+8 \Sigma (l^2) \Sigma [(l-k)^2] l^2 g^{\rho \sigma}
\, .
\end{eqnarray}

By using the loop integral reduction of Passarino and Veltman [Eqs. (\ref{eq:t1}), (\ref{eq:t00}), and (\ref{eq:t11})], Eq. (\ref{eq:scalar_inner_loop}) can be written as
\begin{eqnarray}
i{\cal M}
&\simeq&
-i\frac{N_f \alpha_s (k^2) }{\pi^3 } \epsilon^*_\sigma (k) \epsilon_\rho ( k) g^{\rho \sigma} \delta_{ab} \int d^4 l
\frac{Z[(l-k)^2]\, Z^2 (l^2)}{\Bigl[ (l-k )^2 -\Sigma^2 [(l-k )^2] \Bigr] \Bigl[ l^2 -\Sigma^2 (l^2 ) \Bigr]^2 } 
\nonumber\\
&& \hspace{10em}
\times \Biggl\{
\ S_1(l^2) 
\Biggl[ 
\Bigl[ l^2 + \Sigma^2 (l^2) \Bigr]
\Sigma[(l-k )^2]
\nonumber\\
&& \hspace{15.5em}
+\frac{1}{3k^2 } \Sigma (l^2) \Bigl[ 2k^4 -l^2 k^2 -l^4 + (2l^2 -k^2 )(l-k)^2 -(l -k)^4 \Bigr]
\Biggr]
\nonumber\\
&& \hspace{11.5em}
+S_2(l^2) 
\Biggl[ 
\, 2 l^2 \Sigma (l^2) \Sigma[(l-k )^2]
\nonumber\\
&& \hspace{15.5em}
+\frac{1}{6k^2 } \Bigl[ l^2 + \Sigma^2 (l^2) \Bigr]
\Bigl[ 2k^4 -l^2 k^2 -l^4 + (2l^2 -k^2 )(l-k)^2 -(l -k)^4 \Bigr]
\Biggr]
\ \Biggr\}
\nonumber\\
&\equiv &
\alpha_s (k^2) \epsilon^*_\sigma (k) \epsilon_\rho ( k) g^{\rho \sigma} \delta_{ab} f \bigl[S_1,S_2 ; k^2 \bigr]
\, ,
\end{eqnarray}
where we have omitted terms with the Lorentz structure $k^\rho k^\sigma$, since they cancel when inserted into the second loop (due to the projection of the gluon propagator $g^{\rho \sigma } -k^\rho k^\sigma /k^2$).
Here we have defined the inner loop function $f \bigl[S_1,S_2 ; k^2 \bigr]$.

Now we insert the above inner loop amplitude into the second loop.
The expression of the quark-loop diagram contributing to the SDE (\ref{eq:scalarsde}) is then given by
\begin{eqnarray}
\Gamma^{({\rm QL})} 
&=&
i\frac{C_2 (N_c)}{4\pi^3} \int d^4 k \frac{\alpha_s^2 (k^2)}{k^4} \frac{Z[(p-k)^2 ] f \bigl[S_1,S_2 ; k^2 \bigr] }{(p-k)^2 - \Sigma^2 [(p-k)^2]} 
\left( g^{\rho \sigma} - \frac{k^\rho k^\sigma }{k^2} \right)
\gamma_\sigma \bigl[ p\hspace{-.45em}/\, -k\hspace{-.45em}/\, +\Sigma [(p-k)^2] \bigr] \gamma_\rho 
.
\end{eqnarray}
The Lorentz structure of the above equation is given by
\begin{eqnarray}
\left( g^{\rho \sigma} - \frac{k^\rho k^\sigma }{k^2} \right)
\gamma_\sigma \bigl[ p\hspace{-.45em}/\, -k\hspace{-.45em}/\, +\Sigma [(p-k)^2] \bigr] \gamma_\rho 
&=&
3 k\hspace{-.45em}/\, -p\hspace{-.45em}/\, + 3\Sigma [(p-k)^2] - \frac{1}{k^2} \bigl[ p^2 +k^2 -(p-k)^2 \bigr] k\hspace{-.45em}/\, 
.
\end{eqnarray}

Again by using the Passarino-Veltman reduction [Eqs. (\ref{eq:t1}), (\ref{eq:t00}), and (\ref{eq:t11})] we obtain
\begin{eqnarray}
\Gamma^{({\rm QL})} 
&=&
i\frac{C_2 (N_c)}{4\pi^3} \int d^4 k \, \frac{\alpha_s^2 (k^2)}{ k^4} \frac{Z[(p-k)^2 ] }{ (p-k)^2 - \Sigma^2 [(p-k)^2] } 
f \bigl[S_1,S_2 ; k^2 \bigr] 
\nonumber\\
&& \hspace{6em} \times \Biggl\{ \ 
3 \Sigma [(p-k)^2]
+ \frac{1}{2 p^2 k^2} \Bigl[ 2k^4 -p^2 k^2 -p^4 + (2p^2 -k^2 ) (p-k)^2 -(p-k)^4 \Bigr] p\hspace{-.45em}/\, 
\Biggr\}
.
\end{eqnarray}
After Wick rotation, this gives Eqs. (\ref{eq:sdes1_ql}) and (\ref{eq:sdes2_ql}).

\section{\label{sec:axial_derivation}Detailed calculation of the Schwinger-Dyson equation for the quark axial charge}

\subsection{\label{sec:axial_derivation_1-loop} Quark axial charge SDE: vertex dressing (quenched SDE)}

The Schwinger-Dyson equation for the quark axial charge [Eq. (\ref{eq:axialsde})] is rewritten as
\begin{eqnarray}
&&
G_1 (p^2) \gamma^\mu \gamma_5
+G_2 (p^2) i\sigma^{\mu \nu} p_\nu \gamma_5
+G_3 (p^2) p^\mu p\hspace{-.45em}/\, \gamma_5
\nonumber\\
&=&
\gamma^\mu \gamma_5 +
i C_2 (N_c)
\int \frac{d^4k}{4\pi^3}
\cdot \frac{\alpha_s [(p-k)^2] Z^2 (k^2)}{\left[ k^2 -\Sigma^2 (k^2) \right]^2} 
\cdot \frac{-1}{(p-k)^2} \left[ g^{\rho \lambda} - \frac{(p-k)^\rho (p-k)^\lambda}{(p-k)^2} \right]
\nonumber\\
&& \hspace{7em} \times
\gamma_\rho
\left[ k\hspace{-.45em}/\, + \Sigma (k^2)\right]
\left[ 
G_1 (k^2) \gamma^\mu \gamma_5
+G_2 (k^2) i\sigma^{\mu \nu} k_\nu \gamma_5
+G_3 (k^2) k\hspace{-.45em}/\, \gamma_5 k^\mu
\right]
\left[ k\hspace{-.45em}/\, + \Sigma (k^2)\right] \gamma_\lambda
\, .
\label{eq:axialsde2}
\end{eqnarray}

The Lorentz and Dirac structures of the term with $G_1 (k^2)$ in Eq. (\ref{eq:axialsde2}) can be transformed as
\begin{eqnarray}
&&
\left[ g^{\rho \lambda} - \frac{(p-k)^\rho (p-k)^\lambda}{(p-k)^2} \right]
\gamma_\rho
\left[ k\hspace{-.45em}/\, + \Sigma \right]
\gamma^\mu \gamma_5
\left[ k\hspace{-.45em}/\, + \Sigma \right] 
\gamma_\lambda
\nonumber\\
&=&
\left\{
[k^2 + \Sigma^2 ] \gamma^\mu
+2 \Sigma i \sigma^{\mu \nu} p_\nu  
+2(p\hspace{-.45em}/\, -2k\hspace{-.45em}/\, )k^\mu 
 \right\} 
\gamma_5
\nonumber\\
&&
+\frac{1}{(p-k)^2}
\Biggl\{
2[k^2 + \Sigma^2 ] (p\hspace{-.45em}/\, -k\hspace{-.4em}/\, )
 (p -k )^\mu
-2 \Sigma i \sigma^{\mu \nu} (p -k )_\nu (p^2-k^2) 
\nonumber\\
&& \hspace{10em}
+4 \Sigma  k_\nu i\sigma^{\nu \rho} p_\rho (p -k )^\mu  
-2(p\hspace{-.45em}/\, -k\hspace{-.45em}/\, ) (p^2 -k^2 )k^\mu 
\Biggr\}
\gamma_5
\, .
\label{eq:a1lorentz}
\end{eqnarray}
For simplicity, we have omitted the argument of the self-energy $\Sigma$.
Similarly, the Lorentz and Dirac structures of the term with $G_2 (k^2)$ can be obtained as
\begin{eqnarray}
&&
\left[ g^{\rho \lambda} - \frac{(p-k)^\rho (p-k)^\lambda}{(p-k)^2} \right]
\gamma_\rho
\left[ k\hspace{-.45em}/\, + \Sigma \right]
i\sigma^{\mu \nu} k_\nu \gamma_5
\left[ k\hspace{-.45em}/\, + \Sigma \right] 
\gamma_\lambda
\nonumber\\
&\simeq &
\Biggl\{ 
2 \Sigma [ k^2\gamma^\mu +( p\hspace{-.45em}/\, -2 k\hspace{-.45em}/\, ) k^\mu ]
+\Bigl[ k^2 +\Sigma^2 \Bigr] i\sigma^{\mu \nu} p_\nu
+\frac{ k^2 + \Sigma^2 }{(p-k)^2} \Bigl[ ( k^2 -p^2 ) i\sigma^{\mu \rho} (p-k)_\rho + 2i\sigma^{\nu \rho } k_\nu p_\rho (p-k)^\mu \Bigr]
\nonumber\\
&&\hspace{2em}
+ \frac{2 \Sigma}{(p-k)^2} 
\Bigl[ 2 k^2 p^\mu -(p^2+k^2) k^\mu \Bigr] (p\hspace{-.45em}/\,-k\hspace{-.45em}/\,)
\Biggr\} 
\gamma_5
\, .
\label{eq:a2lorentz}
\end{eqnarray}
The $G_3 (k^2)$ contribution is given as
\begin{eqnarray}
k^\mu
\left[ g^{\rho \lambda} - \frac{(p-k)^\rho (p-k)^\lambda}{(p-k)^2} \right]
\gamma_\rho
\left[ k\hspace{-.45em}/\, + \Sigma \right]
k\hspace{-.45em}/\, \gamma_5 
\left[ k\hspace{-.45em}/\, + \Sigma \right] 
\gamma_\lambda
&=&
k^\mu \left[ k^2-\Sigma^2 \right]
\left\{ 
p\hspace{-.45em}/\,-2k\hspace{-.45em}/\, 
+ \frac{k^2-p^2}{(p-k)^2} 
(p\hspace{-.45em}/\,- k\hspace{-.45em}/\,)
\right\}
\gamma_5
\, .
\label{eq:a3lorentz}
\end{eqnarray}

By substituting Eqs. (\ref{eq:a1lorentz}), (\ref{eq:a2lorentz}), and (\ref{eq:a3lorentz}) into Eq. (\ref{eq:axialsde2}), we can further transform the integral equation as
\begin{eqnarray}
&&
G_1 (p^2) \gamma^\mu \gamma_5
+G_2 (p^2) i\sigma^{\mu \nu} p_\nu \gamma_5
+G_3 (p^2) p\hspace{-.45em}/\, \gamma_5 p^\mu
\nonumber\\
&=&
\gamma^\mu \gamma_5 
\nonumber\\
&&
-i \int \frac{d^4k}{4\pi^3}
\cdot \frac{\alpha_s [(p-k)^2] Z^2 (k^2)}{\left[ k^2 -\Sigma^2 (k^2) \right]^2} 
\cdot \frac{C_2 (N_c)}{(p-k)^2} \cdot G_1(k^2)
\nonumber\\
&& \hspace{5em} \times 
\Biggl\{
\frac{1}{3} \gamma^\mu \gamma_5 \Biggl[ 
-\frac{p^2+ \Sigma^2 (k^2)}{2p^2} \frac{(p^2-k^2)^2}{(p-k)^2}
+\frac{2p^4 +2p^2 k^2+k^4 +\Sigma^2 (k^2) \left( 4p^2 +k^2 \right)}{p^2} 
\nonumber\\
&& \hspace{22em} 
-\frac{5p^2+4k^2 +\Sigma^2 (k^2)}{2p^2} (p-k)^2
+\frac{(p-k)^4}{p^2}
\Biggr]
\nonumber\\
&& \hspace{6em}
+
\frac{1}{3 p^2}
\Sigma (k^2)  i\sigma^{\mu \nu} p_\nu \gamma_5
\left[
- 2\frac{(p^2-k^2)^2}{(p-k)^2} 
+p^2+ k^2
+(p-k)^2
\right]
\nonumber\\
&& \hspace{6em}
+\frac{1}{3} p^\mu p\hspace{-.45em}/\, \gamma_5 
\Biggl[ 
\frac{2\Sigma^2 (k^2) -p^2}{p^4} \frac{(p^2 -k^2)^2}{(p-k)^2}
-\frac{2p^4+2p^2 k^2 +4k^4 - 2(p^2 -2k^2) \Sigma^2 ( k^2)}{p^4}
\nonumber\\
&& \hspace{22em} 
+ \frac{7p^2 +8k^2 +2\Sigma^2(k^2)}{p^4} (p-k)^2
-4(p-k)^4
\Biggr]
\ \Biggr\}
\nonumber\\
&&
-i \int \frac{d^4k}{4\pi^3}
\cdot \frac{\alpha_s [(p-k)^2] Z^2 (k^2)}{\left[ k^2 -\Sigma^2 (k^2) \right]^2} 
\cdot \frac{C_2 (N_c)}{(p-k)^2} \cdot G_2(k^2)
\nonumber\\
&& \hspace{5em} \times 
\Biggl\{
\frac{-1}{3p^2} \Sigma (k^2) \gamma^\mu \gamma_5 
\left[ 
\frac{p^2 +k^2}{2} \frac{(p^2-k^2)^2}{(p-k)^2} 
-2(p^4 +3 p^2 k^2+k^4)
+\frac{5}{2} (p^2+k^2) (p-k)^2
-(p-k)^4
\right]
\nonumber\\
&& \hspace{6.2em}
+
\frac{1}{6p^2} \left[ k^2 + \Sigma^2 (k^2) \right] 
i\sigma^{\mu \nu} p_\nu \gamma_5
\left[
-2 \frac{(p^2-k^2)^2}{(p-k)^2}
+p^2+k^2
+(p-k)^2
\right]
\nonumber\\
&& \hspace{6.2em}
-\frac{1}{3p^4} \Sigma (k^2) p^\mu p\hspace{-.45em}/\, \gamma_5 
\left[
(p^2 -2k^2) \frac{(p^2- k^2)^2}{(p-k)^2}
+2p^4 +8k^4
-(7p^2+10k^2) (p-k)^2
+4(p-k)^4
\right]
\ \Biggr\}
\nonumber\\
&&
-i \int \frac{d^4k}{4\pi^3}
\cdot \frac{\alpha_s [(p-k)^2] Z^2 (k^2)}{ k^2 -\Sigma^2 (k^2)} 
\cdot \frac{C_2 (N_c)}{(p-k)^2} \cdot G_3(k^2)
\cdot \left[ k^2 -\Sigma^2 (k^2) \right] 
\nonumber\\
&& \hspace{5em} \times 
\Biggl\{
\frac{1}{6p^2} \gamma^\mu \gamma_5 
\left[
-\frac{1}{2} \frac{(p^2-k^2)^3}{(p-k)^2} 
+2p^2 (p^2-k^2) 
-\frac{1}{2} (5p^2+3k^2) (p-k)^2
+(p-k)^4
\right]
\nonumber\\
&&\hspace{6.2em}
+\frac{1}{6p^4} p^\mu p\hspace{-.45em}/\, \gamma_5 
\left[
-(p^2+2k^2) \frac{(p^2-k^2)^2}{(p-k)^2}
-2p^2(p^2 +2k^2 ) 
+(7p^2+6k^2)(p-k)^2
-4(p-k)^4
\right]
 \Biggr\}
.
\label{eq:axialsde3}
\end{eqnarray}
As for the calculation of the scalar density SDE, we have used the formulae (\ref{eq:t1}), (\ref{eq:t00}), and (\ref{eq:t11})  to reduce into a Lorentz scalar loop integral.

By taking the trace after multiplying by $\gamma_\mu \gamma_5$, Eq. (\ref{eq:axialsde3}) can be rewritten as
\begin{eqnarray}
4 G_1 (p^2) +G_3 (p^2) p^2
&=&
4
-i \frac{C_2 (N_c)}{4\pi^3} \int d^4 k \, \frac{\alpha_s \left[ (p-k)^2 \right] Z^2 (k^2)}{\left[ k^2 - \Sigma^2 (k^2) \right]^2}
\nonumber\\
&& \hspace{6em} \times \Biggl\{ \ \ \ \,
G_1 (k^2) \left[ -\frac{(p^2 -k^2)^2}{(p-k)^4} +2 \frac{p^2+k^2 +3\Sigma^2 (k^2)}{(p-k)^2} -1 \right]
\nonumber\\
&& \hspace{8em} 
+G_2 (k^2) \Sigma (k^2) \left[ -\frac{(p^2 -k^2)^2}{(p-k)^4} +2 \frac{p^2+4k^2 }{(p-k)^2} -1 \right]
\nonumber\\
&& \hspace{8em} 
+G_3 (k^2) \frac{k^2- \Sigma^2 (k^2)}{2} \left[ -\frac{(p^2 -k^2)^2}{(p-k)^4} +2 \frac{p^2-2k^2 }{(p-k)^2} -1 \right]
\ \Biggr\} 
\ .
\label{eq:g1g3_1}
\end{eqnarray}
In the derivation of the above equation, we have used
\begin{eqnarray}
{\rm Tr} \left[ \gamma^\mu \gamma_5 \gamma_\mu \gamma_5 \right]
&=&
-16
\, ,
\label{eq:traceformula1}
\\
{\rm Tr} \left[ p^\mu p\hspace{-0.5em}/\, \gamma_5 \gamma_\mu \gamma_5 \right]
&=&
-4p^2
\, .
\label{eq:traceformula2}
\end{eqnarray}

By taking the trace after multiplying by $i\sigma_{\mu \rho} p^\rho \gamma_5$, Eq. (\ref{eq:axialsde3}) can be rewritten as
\begin{eqnarray}
G_2 (p^2)
&=&
-i \frac{C_2 (N_c)}{12\pi^3} \int d^4 k \, \frac{\alpha_s \left[ (p-k)^2 \right] Z^2 (k^2)}{\left[ k^2 - \Sigma^2 (k^2) \right]^2}
\cdot \frac{1}{p^2} \left[ 
-2 \frac{(p^2 -k^2)^2}{(p-k)^4} + \frac{p^2+k^2 }{(p-k)^2} +1
\right]
\nonumber\\
&& \hspace{16em} \times \Biggl\{ \ 
G_1 (k^2) \Sigma (k^2) 
+G_2 (k^2) \frac{k^2 +\Sigma^2 (k^2)}{2}
\ \Biggr\} 
\ .
\label{eq:g2}
\end{eqnarray}
In the derivation of the above equation, we have used
\begin{eqnarray}
{\rm Tr} \left[ \gamma^\mu \gamma_5 i\sigma_{\mu \rho} p^\rho \gamma_5 \right]
&=&
0
\, ,
\\
{\rm Tr} \left[ i\sigma_{\mu \nu} p^\nu \gamma_5 i\sigma^{\mu \rho} p_\rho \gamma_5 \right]
&=&
-12p^2
\, ,
\\
{\rm Tr} \left[ p^\mu p\hspace{-0.5em}/\, \gamma_5 i\sigma_{\mu \rho} p^\rho \gamma_5 \right]
&=&
0
\, .
\end{eqnarray}

By taking the trace after multiplying by $p_\mu p\hspace{-0.5em}/\, \gamma_5$, Eq. (\ref{eq:axialsde2}) can be rewritten as
\begin{eqnarray}
G_1 (p^2) +G_3 (p^2) p^2
&=&
1
-i \frac{C_2 (N_c)}{4\pi^3} \int d^4 k \, \frac{\alpha_s \left[ (p-k)^2 \right] Z^2 (k^2)}{\left[ k^2 - \Sigma^2 (k^2) \right]^2}
\nonumber\\
&& \hspace{6em} \times \Biggl\{ \ 
G_1 (k^2) \Biggl[ 
\frac{1}{2} \left( \frac{\Sigma^2 (k^2) }{p^2} -1 \right) \frac{(p^2 -k^2)^2}{(p-k)^4} 
+ \frac{-k^4 +\Sigma^2 (k^2) (2p^2 -k^2 ) }{ p^2 (p-k)^2} 
\nonumber\\
&& \hspace{20em}
+\frac{3}{2} +2 \frac{k^2}{p^2} +\frac{1}{2} \frac{ \Sigma^2(k^2) }{p^2} 
-\frac{(p-k)^2}{p^2}
\Biggr]
\nonumber\\
&& \hspace{7em} 
+G_2 (k^2) \Sigma (k^2) \Biggl[ 
-\frac{p^2 -k^2}{2p^2} \frac{(p^2 -k^2)^2}{(p-k)^4} 
-2 \frac{k^4-p^2 k^2 }{p^2 (p-k)^2} 
\nonumber\\
&& \hspace{22em}
+\frac{3p^2 +5k^2 }{2p^2} 
-\frac{(p-k)^2}{p^2}
\Biggr]
\nonumber\\
&& \hspace{7em} 
+G_3 (k^2) \frac{k^2- \Sigma^2 (k^2)}{2} \Biggl[ 
-\frac{p^2+k^2}{2p^2} \frac{(p^2 -k^2)^2}{(p-k)^4} 
-2 \frac{k^2 }{(p-k)^2} 
\nonumber\\
&& \hspace{22em}
+\frac{3}{2} \frac{p^2+k^2}{p^2}
-\frac{(p-k)^2}{p^2}
 \Biggr]
\ \Biggr\} 
\ .
\label{eq:g1g3_2}
\end{eqnarray}
In the derivation of the above equation, we have used Eq. (\ref{eq:traceformula2}) and 
\begin{eqnarray}
{\rm Tr} \left[ p^\mu p\hspace{-0.5em}/\, \gamma_5 p_\mu p\hspace{-0.5em}/\, \gamma_5 \right]
&=&
-4p^4
\, .
\label{eq:traceformula3}
\end{eqnarray}

By equating Eqs. (\ref{eq:g1g3_1}) and (\ref{eq:g1g3_2}), we obtain the system of integral equations for $G_1$, $G_2$, and $G_3$,
\begin{eqnarray}
G_1 (p^2) 
&=&
1
-i \frac{C_2 (N_c)}{12\pi^3} \int d^4 k \, \frac{\alpha_s \left[ (p-k)^2 \right] Z^2 (k^2)}{\left[ k^2 - \Sigma^2 (k^2) \right]^2 p^2}
\nonumber\\
&& \hspace{6em} \times \Biggl\{ \ 
G_1 (k^2) \Biggl[ 
-\frac{\Sigma^2 (k^2) +p^2}{2} \frac{(p^2 -k^2)^2}{(p-k)^4} 
+ \frac{2p^4 +2p^2k^2 +k^4 +\Sigma^2 (k^2) (4p^2 + k^2 ) }{ (p-k)^2} 
\nonumber\\
&& \hspace{26em}
-\frac{ 5p^2+4k^2+\Sigma^2 (k^2) }{2} 
+(p-k)^2
\Biggr]
\nonumber\\
&& \hspace{7em} 
+G_2 (k^2) \Sigma (k^2) \Biggl[ 
-\frac{p^2 +k^2}{2} \frac{(p^2 -k^2)^2}{(p-k)^4} 
+ 2\frac{p^4 +3p^2 k^2 + k^4 }{ (p-k)^2} 
\nonumber\\
&& \hspace{24em}
-\frac{5}{2} (p^2 +k^2 )
+(p-k)^2
\Biggr]
\nonumber\\
&& \hspace{7em} 
+G_3 (k^2) \frac{k^2- \Sigma^2 (k^2)}{2} \Biggl[ 
\frac{k^2-p^2}{2} \frac{(p^2 -k^2)^2}{(p-k)^4} 
+2 \frac{p^2 (p^2- k^2 )}{(p-k)^2} 
\nonumber\\
&& \hspace{24em}
-\frac{5p^2+3k^2}{2}
+(p-k)^2
 \Biggr]
\ \Biggr\} 
\ ,
\label{eq:sdeg1'}
\\
G_2 (p^2)
&=&
-i\frac{C_2 (N_c)}{12\pi^3} \int d^4 k \, \frac{\alpha_s \left[ (p-k)^2 \right] Z^2 (k^2)}{\left[ k^2 - \Sigma^2 (k^2) \right]^2 p^2}
\left[ 
-2 \frac{(p^2 -k^2)^2}{(p-k)^4} + \frac{p^2+k^2 }{(p-k)^2} +1
\right]
\nonumber\\
&& \hspace{20em} \times \Biggl\{ \ 
G_1 (k^2) \Sigma (k^2) 
+G_2 (k^2) \frac{k^2 +\Sigma^2 (k^2)}{2}
\ \Biggr\} 
\ ,
\\
G_3 (p^2) 
&=&
-i\frac{C_2 (N_c)}{12\pi^3} \int d^4 k \, \frac{\alpha_s \left[ (p-k)^2 \right] Z^2 (k^2)}{\left[ k^2 - \Sigma^2 (k^2) \right]^2 p^4}
\nonumber\\
&& \hspace{6em} \times \Biggl\{ \ 
G_1 (k^2) \Biggl[ 
\left[ 2\Sigma^2 (k^2) -p^2 \right] \frac{(p^2 -k^2)^2}{(p-k)^4} 
-2 \frac{p^4 +p^2k^2 +2k^4 +\Sigma^2 (k^2) (2 k^2-p^2 ) }{ (p-k)^2} 
\nonumber\\
&& \hspace{24em}
+7 p^2 +8 k^2 +2\Sigma^2 (k^2)
-4 (p-k)^2
\Biggr]
\nonumber\\
&& \hspace{7em} 
+G_2 (k^2) \Sigma (k^2) \Biggl[ 
-(p^2-2k^2)\frac{(p^2 -k^2)^2}{(p-k)^4} 
- 2\frac{p^4 + 4 k^4 }{ (p-k)^2} 
+7p^2 +10k^2
-4(p-k)^2
\Biggr]
\nonumber\\
&& \hspace{7em} 
+G_3 (k^2) \frac{k^2- \Sigma^2 (k^2)}{2} \Biggl[ 
-(p^2+2k^2) \frac{(p^2 -k^2)^2}{(p-k)^4} 
-2 \frac{p^2(p^2 + 2k^2 )}{(p-k)^2} 
\nonumber\\
&& \hspace{26em}
+ 7p^2+6k^2
-4(p-k)^2
 \Biggr]
\ \Biggr\} 
\ .
\label{eq:sdeg3'}
\end{eqnarray}
By Wick rotating the above equations, we obtain Eqs. (\ref{eq:sdeg1}), (\ref{eq:sdeg2}), and (\ref{eq:sdeg3}).

\subsection{\label{sec:axial_derivation_2-loop} Quark axial charge SDE: quark-loop contribution (unquenched isoscalar SDE)}

To obtain the analytical expression of the quark-loop diagram contribution of the SDE (\ref{eq:axialsde}) (the last term of the right-hand side of the SDE  of Fig. \ref{fig:axial_SD_eq}), we first calculate the inner quark-loop integral (see Fig. \ref{fig:axial_inner_loop}).

The amplitude of the inner quark-loop is given as
\begin{eqnarray}
i{\cal M}_5^\mu
&=&
-iN_f \frac{\alpha_s (k^2)}{4\pi^3} 
\int d^4 l
\frac{{\rm Tr} \left[ \left\{ l\hspace{-.45em}/\, -2k\hspace{-.45em}/\, +\Sigma [(l-2k)^2] \right\} \gamma^\sigma \left\{ l\hspace{-.45em}/\, -k\hspace{-.45em}/\, +\Sigma [(l-k)^2] \right\} \Gamma_5^\mu (l) \left\{ l\hspace{-.45em}/\, -k\hspace{-.45em}/\, +\Sigma [(l-k)^2] \right\} \gamma^\rho \right] }{ \left\{ (l-2k)^2 -\Sigma^2 [(l-2k)^2] \right\} \left\{ (l-k)^2 - \Sigma^2 [(l-k)^2] \right\}^2 }
\nonumber\\
&& \hspace{10em} \times
Z [(l-2k)^2] Z^2 [(l-k)^2]
\epsilon^*_\sigma (k) \epsilon_\rho ( k)
\delta_{ab}
\, ,
\label{eq:axial_inner_loop}
\end{eqnarray}
where $\Gamma_5^\mu (l) \equiv G_1 (l^2) \gamma^\mu \gamma_5
+G_2 (l^2) i\sigma^{\mu \nu} l_\nu \gamma_5
+G_3 (l^2) l^\mu l\hspace{-.45em}/\, \gamma_5 
$.
The color trace has already been reduced (${\rm tr}[t_a t_b ] = \frac{1}{2} \delta_{ab}$).
Here we have included the contribution from the loop with the oppositely propagating quark, similarly as for the scalar quark, since the axial vector currents of the quark and antiquark do not change the sign [see the discussion below Eq. (\ref{eq:scalar_inner_loop})].

Let us first consider the $G_1 $ contribution.
The trace is calculated as
\begin{eqnarray}
&&
{\rm Tr} \left[ \left\{ l\hspace{-.45em}/\, -2k\hspace{-.45em}/\, +\Sigma [(l-2k)^2] \right\} \gamma^\sigma \left\{ l\hspace{-.45em}/\, -k\hspace{-.45em}/\, +\Sigma [(l-k)^2] \right\} \gamma^\mu \gamma_5\left\{ l\hspace{-.45em}/\, -k\hspace{-.45em}/\,+\Sigma [(l-k)^2] \right\} \gamma^\rho \right] 
\nonumber\\
&=&
4i\epsilon^{\mu \alpha \sigma \rho}
\left[ \left\{ (l-k)^2 + \Sigma^2 [(l-k)^2] \right\} (l_\alpha -2k_\alpha ) - 2 \Sigma [(l-2k)^2] \Sigma [(l-k)^2] \, (l_\alpha -k_\alpha ) \right]
+8i (l^\mu -k^\mu ) \epsilon^{\beta \alpha \sigma \rho} l_\beta k_\alpha
, \ \ \ \ \ \ 
\end{eqnarray}
where we have used ${\rm Tr} [\gamma^\mu \gamma^\nu \gamma^\rho \gamma^\sigma \gamma_5] =-4i \epsilon^{\mu \nu \rho \sigma}$.

The $G_2$ contribution is calculated similarly as
\begin{eqnarray}
&&
{\rm Tr} \left[ \left\{ l\hspace{-.45em}/\, -2k\hspace{-.45em}/\, +\Sigma [(l-2k)^2] \right\} \gamma^\sigma \left\{ l\hspace{-.45em}/\, -k\hspace{-.45em}/\, +\Sigma [(l-k)^2] \right\} i\sigma^{\mu \nu} ( l_\nu -k_\nu ) \gamma_5 \left\{ l\hspace{-.45em}/\, -k\hspace{-.45em}/\,+\Sigma [(l-k)^2] \right\} \gamma^\rho \right] 
\nonumber\\
&=&
4i\epsilon^{\mu \alpha \sigma \rho}
\left[ 2 (l-k)^2 \Sigma [(l-k)^2]  (l_\alpha -2k_\alpha ) - \left\{ (l-k)^2 + \Sigma^2 [(l-k)^2] \right\} \Sigma [(l-2k)^2] \, (l_\alpha -k_\alpha) \right]
\nonumber\\
&&
+8 \Sigma [(l-k)^2] (l^\mu -k^\mu ) i\epsilon^{\beta \alpha \sigma \rho} l_\beta k_\alpha
\, .
\end{eqnarray}

The Dirac trace of the $G_3$ contribution is given as
\begin{eqnarray}
&&
{\rm Tr} \left[ \left\{ l\hspace{-.45em}/\, -2k\hspace{-.45em}/\, +\Sigma [(l-2k)^2] \right\} \gamma^\sigma \left\{ l\hspace{-.45em}/\, -k\hspace{-.45em}/\, +\Sigma [(l-k)^2] \right\} ( l_\mu -k_\mu ) (l\hspace{-.45em}/\, - k\hspace{-.45em}/\, ) \gamma_5 \left\{ l\hspace{-.45em}/\, -k\hspace{-.45em}/\,+\Sigma [(l-k)^2] \right\} \gamma^\rho \right] 
\nonumber\\
&=&
4i \left\{ (l-k)^2 - \Sigma^2 [(l-k)^2] \right\}
( l^\mu -k^\mu ) \epsilon^{\beta \alpha \sigma \rho} l_\beta k_\alpha
\, .
\end{eqnarray}

By using the loop integral reduction of Passarino and Veltman [Eqs. (\ref{eq:t1}), (\ref{eq:t00}) and (\ref{eq:t11})], Eq. (\ref{eq:axial_inner_loop}) can be written as
\begin{eqnarray}
i{\cal M}_5^\mu
&=&
\frac{N_f \alpha_s (k^2) }{2\pi^3 k^2} \int d^4 l
\frac{Z^2 [(l-k)^2]\, Z[(l-2k)^2 ] \epsilon^*_\sigma (k) \epsilon_\rho ( k) \epsilon^{\mu \alpha \sigma \rho} k_\alpha \delta_{ab}}{\Bigl[ (l-2k )^2 -\Sigma^2 [(l-2k )^2] \Bigr] \Bigl[ (l-k)^2 -\Sigma^2 [(l-k)^2] \Bigr]^2 } 
\nonumber\\
&& \hspace{6em}
\times \Biggl\{
\ G_1 [(l-k)^2] 
\Biggl[ 
\frac{1}{3} \Bigl[ -(l^2 -k^2)^2 +(5l^2 -7k^2 )(l-k)^2 -4(l -k)^4 \Bigr]
\nonumber\\
&& \hspace{14em}
+\Sigma^2 [(l-k)^2]
\Bigl[ l^2 -3k^2 - (l-k)^2 \Bigr]
\nonumber\\
&& \hspace{14em}
+2\Sigma \bigl[ (l-2k)^2 \bigr] \Sigma [(l-k)^2] \Bigl[ k^2 - l^2 +(l-k)^2 \Bigr]
\Biggr]
\nonumber\\
&& \hspace{7.5em}
+G_2 [(l-k)^2]
\Biggl[ 
\frac{1}{3} \Sigma [(l-k)^2] 
\Bigl[
-(l^2 -k^2)^2 +8(l^2-2k^2)(l-k)^2-7(l-k)^4
\Bigr]
\nonumber\\
&& \hspace{14em}
+\Sigma \bigl[ (l-2k)^2 \bigr] 
\Bigl[ (l-k)^2 + \Sigma^2 [(l-k)^2] \Bigr]
\Bigl[ k^2 -l^2 + (l-k)^2 \Bigr]
\Biggr]
\nonumber\\
&& \hspace{7.5em}
+G_3 [(l-k)^2]
\frac{1}{6} 
\Bigl[ (l-k)^2 - \Sigma^2 [(l-k)^2] \Bigr]
\Bigl[
-(l^2 -k^2)^2 +2(l^2 +k^2)(l-k)^2 -(l -k)^4
\Bigr]
\ \Biggr\}
\nonumber\\
&\equiv &
\alpha_s (k^2)  \epsilon^*_\sigma (k) \epsilon_\rho ( k) i\epsilon^{\mu \alpha \sigma \rho} k_\alpha \delta_{ab} f_5 \bigl[G_1,G_2,G_3 ; k^2 \bigr]
\, .
\end{eqnarray}
Here we have defined the inner loop function $f_5 \bigl[G_1,G_2,G_3 ; k^2 \bigr]$.

Now we insert the above inner loop amplitude into the second loop.
The expression of the quark-loop diagram contributing to the SDE (\ref{eq:axialsde}) is then given by
\begin{eqnarray}
\Gamma_{5}^{({\rm QL})\mu} 
&=&
-\frac{C_2 (N_c)}{4\pi^3} \int d^4 k \frac{\alpha_s^2 (k^2)}{k^4} \frac{Z[(p-k)^2 ] f_5 \bigl[G_1,G_2,G_3 ; k^2 \bigr] }{(p-k)^2 - \Sigma^2 [(p-k)^2]} 
\gamma_\sigma \bigl[ p\hspace{-.45em}/\, -k\hspace{-.45em}/\, +\Sigma [(p-k)^2] \bigr] \gamma_\rho \epsilon^{\mu \alpha \sigma \rho} k_\alpha
.
\end{eqnarray}
The Lorentz structure of the above equation is given by
\begin{eqnarray}
\gamma_\sigma \bigl[ p\hspace{-.45em}/\, -k\hspace{-.45em}/\, +\Sigma [(p-k)^2] \bigr] \gamma_\rho \epsilon^{\mu \alpha \sigma \rho} k_\alpha
&=&
2\sigma^{\mu \alpha} k_\alpha \gamma_5 \bigl[ p\hspace{-.45em}/\, -k\hspace{-.45em}/\, -\Sigma [(p-k)^2] \bigr]
+2\gamma_\sigma \epsilon^{\mu \alpha \sigma \rho} k_\alpha p_\rho
,
\end{eqnarray}
where we have used $\epsilon^{\mu \alpha \sigma \rho} \sigma_{\rho \sigma} = -2i \gamma_5 \sigma^{\mu \alpha}$.
The last term $\gamma_\rho \epsilon^{\mu \alpha \sigma \rho} k_\alpha p_\rho$ of the above equation does not contribute to the final result, since it cancels after the Passarino-Veltman reduction [see Eq. (\ref{eq:t1})].

Again, by using the Passarino-Veltman reduction [Eqs. (\ref{eq:t1}), (\ref{eq:t00}), and (\ref{eq:t11})] we obtain
\begin{eqnarray}
\Gamma_{5}^{({\rm QL})\mu} 
&=&
-i\frac{C_2 (N_c)}{4\pi^3} \int d^4 k \frac{\alpha_s^2 (k^2)}{p^2 k^4} \frac{Z[(p-k)^2 ] f_5 \bigl[G_1,G_2,G_3 ; k^2 \bigr] }{ (p-k)^2 - \Sigma^2 [(p-k)^2] } 
\nonumber\\
&& \hspace{6em} \times \Biggl\{ \ 
\frac{1}{6} \Bigl[ 
-5p^4 +4p^2 k^2 +k^4 +(4p^2 -2k^2) (p-k)^2 +(p-k)^4
\Bigr] \gamma^\mu \gamma_5
\nonumber\\
&& \hspace{8em} 
+ \Sigma [(p-k)^2] \Bigl[ 
p^2+k^2-(p-k)^2
\Bigr] i\sigma^{\mu \nu} p_\nu \gamma_5
\nonumber\\
&& \hspace{8em} 
+\frac{1}{3} \Bigl[ 
p^4 +p^2 k^2 -2k^4 +(p^2+4k^2)(p-k)^2 -2(p-k)^4
\Bigr] 
\frac{p^\mu p\hspace{-.45em}/\, \gamma_5 }{p^2}
\Biggr\}
.
\end{eqnarray}
The SDE for the quark isoscalar axial charge is given by adding $\Gamma_{5}^{({\rm QL})\mu} $ to Eq. (\ref{eq:axialsde3}).
The isoscalar axial SDE [Eq. (\ref{eq:axialsde3}) augmented with $\Gamma_{5}^{({\rm QL})\mu} $] for the $G_2$ function is easily obtained by taking the trace with $i\sigma^{\mu \nu} p_\nu \gamma_5$. 
By Wick rotating it, we have Eq. (\ref{eq:sdeg2_isoscalar}).

To obtain the contribution of $\Gamma_{5}^{({\rm QL})\mu} $ to $G_1 (p^2)$ and $G_3 (p^2)$, we must equate $G_1 (p^2)$ and $G_3 (p^2)$ after taking the trace of the isoscalar axial SDE (SDE with $\Gamma_{5}^{({\rm QL})\mu} $) with $\gamma^\mu \gamma_5$ and $p^\mu p\hspace{-.45em}/\, \gamma_5$.
The trace of the isoscalar axial SDE with $\gamma^\mu \gamma_5$ gives
\begin{equation}
4 G_1 (p^2) + G_3 (p^2) p^2
=
-i\frac{3 C_2 (N_c)}{4\pi^3} \int d^4 k \frac{\alpha_s^2 (k^2)}{ k^4} \frac{Z[(p-k)^2 ] f_5 \bigl[G_1,G_2,G_3 ; k^2 \bigr] }{ (p-k)^2 - \Sigma^2 [(p-k)^2] } 
\Bigl[ 
k^2 -p^2 + (p-k)^2 
\Bigr] 
+ [\mbox{RHS of Eq. (\ref{eq:g1g3_1})} ]
,
\end{equation}
where we have used Eqs. (\ref{eq:traceformula1}) and (\ref{eq:traceformula2}).
On the other hand, the trace of the isoscalar axial SDE with $p^\mu p\hspace{-.45em}/\, \gamma_5$ yields
\begin{eqnarray}
G_1 (p^2) + G_3 (p^2) p^2
&=&
-i\frac{C_2 (N_c)}{8\pi^3} \int d^4 k \frac{\alpha_s^2 (k^2)}{p^2 k^4} \frac{Z[(p-k)^2 ] f_5 \bigl[G_1,G_2,G_3 ; k^2 \bigr] }{ (p-k)^2 - \Sigma^2 [(p-k)^2] } 
\nonumber\\
&& \hspace{6em} \times
\Bigl[ 
-(p^2-k^2)^2 +2(p^2 +k^2) (p-k)^2 -(p-k)^4
\Bigr] 
+ [\mbox{RHS of Eq. (\ref{eq:g1g3_2})} ]
,\ \ \ \ \ 
\end{eqnarray}
where we have used Eqs. (\ref{eq:traceformula2}) and (\ref{eq:traceformula3}).

By equating the above two equations, we obtain
\begin{eqnarray}
G_1 (p^2) 
&=&
-i\frac{C_2 (N_c)}{12\pi^3} \int d^4 k \frac{\alpha_s^2 (k^2)}{p^2 k^4} \frac{Z[(p-k)^2 ] f_5 \bigl[G_1,G_2,G_3 ; k^2 \bigr] }{ (p-k)^2 - \Sigma^2 [(p-k)^2] } 
\nonumber\\
&& \hspace{6em} \times
\Biggl[ 
-\frac{5}{2} p^4 +2p^2 k^2 +\frac{1}{2} k^4 +(2p^2-k^2) (p-k)^2 +\frac{1}{2}(p-k)^4
\Biggr] 
+ [\mbox{RHS of Eq. (\ref{eq:sdeg1'})} ]
, \ \ \ \ 
\\
G_3 (p^2)
&=&
-i\frac{C_2 (N_c)}{12\pi^3} \int d^4 k \frac{\alpha_s^2 (k^2)}{p^4 k^4} \frac{Z[(p-k)^2 ] f_5 \bigl[G_1,G_2,G_3 ; k^2 \bigr] }{ (p-k)^2 - \Sigma^2 [(p-k)^2] } 
\nonumber\\
&& \hspace{6em} \times
\Biggl[ 
p^4 +p^2 k^2 - 2 k^4 +(p^2+4k^2) (p-k)^2 -2 (p-k)^4
\Biggr] 
+ [\mbox{RHS of Eq. (\ref{eq:sdeg3'})} ]
. \ \ \ \ 
\end{eqnarray}
After Wick rotation, we obtain Eqs. (\ref{eq:sdeg1_isoscalar}) and (\ref{eq:sdeg3_isoscalar}).

\twocolumngrid

\end{document}